\documentclass[%
 reprint,
superscriptaddress,
 amsmath,amssymb,
 aps,
prx,
floatfix,
]{revtex4-2}

\usepackage{graphicx}
\usepackage{dcolumn}
\usepackage{bm}
\usepackage{physics}
\usepackage{tikz}
\usepackage{quantikz}
\usepackage{amsmath}
\usepackage{xcolor}
\usepackage{adjustbox}
\definecolor{qft}{RGB}{56,134,210}
\definecolor{phase}{RGB}{220,140,40}
\definecolor{dil}{RGB}{180,60,60}
\definecolor{nh}{RGB}{60,160,90}

\usepackage[colorlinks=true,allcolors=blue]{hyperref}
\usepackage{orcidlink}

\newcommand{\ii}{\mathrm{i}}

\newcommand{\NVE}{\mathrm{NVE}}
\newcommand{\NVT}{\mathrm{NVT}}
\newcommand{\QPE}{\mathrm{QPE}}

\newcommand{\Hkvn}{\hat H_{\mathrm{KvN}}}
\newcommand{\Hthree}{\hat H_3}
\newcommand{\Dbart}{D_{\mathrm{Bart}}}

\newcommand{\Cvv}{C_{vv}}

\newcommand{\bigO}{\mathcal{O}}

\begin{document}
\title{Koopman--von Neumann Molecular Dynamics for Green--Kubo Transport Coefficients}
\author{Masari Watanabe\orcidlink{0000-0003-3186-535X}}
\email{mwatanabe@quemix.com}
\affiliation{Quemix Inc., Taiyo Life Nihonbashi Building, 2-11-2,
Nihonbashi Chuo-ku, Tokyo 103-0027, Japan}
\affiliation{Department of Physics, The University of Tokyo, Hongo, Bunkyo-ku, Tokyo 113-0033, Japan}

\author{Hirofumi Nishi\orcidlink{0000-0001-5155-6605}}
\affiliation{Quemix Inc., Taiyo Life Nihonbashi Building, 2-11-2,
Nihonbashi Chuo-ku, Tokyo 103-0027, Japan}
\affiliation{Department of Physics, The University of Tokyo, Hongo, Bunkyo-ku, Tokyo 113-0033, Japan}

\author{Taichi Kosugi\orcidlink{0000-0003-3379-3361}}
\affiliation{Quemix Inc., Taiyo Life Nihonbashi Building, 2-11-2,
Nihonbashi Chuo-ku, Tokyo 103-0027, Japan}
\affiliation{Department of Physics, The University of Tokyo, Hongo, Bunkyo-ku, Tokyo 113-0033, Japan}

\author{Shigekazu Hidaka\orcidlink{0000-0002-7992-1018}}
\affiliation{Advanced Research and Innovation Center, DENSO CORPORATION, 500-1 Minamiyama, Komenoki-cho, Nisshin, Aichi 470-0111, Japan}

\author{Ryo Sakurai\orcidlink{0009-0004-2721-7231}}
\affiliation{Advanced Research and Innovation Center, DENSO CORPORATION, 1-1-4, Haneda Airport, Ota-ku, Tokyo 144-0041, Japan}

\author{Yu-ichiro Matsushita\orcidlink{0000-0002-9254-5918}}
\affiliation{Quemix Inc., Taiyo Life Nihonbashi Building, 2-11-2, Nihonbashi Chuo-ku, Tokyo 103-0027, Japan}
\affiliation{Department of Physics, The University of Tokyo, Hongo, Bunkyo-ku, Tokyo 113-0033, Japan}
\affiliation{Quantum Materials and Applications Research Center,
National Institutes for Quantum Science and Technology, Tokyo 152-8550, Japan}

\date{\today}

\begin{abstract}
We formulate the Green--Kubo transport coefficients of classical molecular dynamics as a readout problem for quantum algorithms using the Koopman--von Neumann (KvN) representation. 
Both NVE and Nos\'e--Hoover-type NVT dynamics are derived as unitary evolutions on Hilbert spaces associated with the corresponding classical phase spaces.
Numerical benchmarks on finite grids show that the discretization error in the correlation function decreases as a power law in the number of grid points $N_z$.
Equivalently, with $N_z=2^{n_z}$, the error decreases exponentially in the register size $n_z$, so a target accuracy $\epsilon$ requires $n_z=\bigO(\log(1/\epsilon))$ qubits. To read out a transport coefficient, we input a flux-excited state to quantum phase estimation (QPE). 
The probability $P_0$ of measuring the QPE ancilla register in the all-zero state corresponds to a Bartlett-windowed Green--Kubo integral.
With maximum-likelihood amplitude estimation, the statistical estimation of $P_0$ defined by this QPE oracle improves from the $N_{\rm queries}^{-1/2}$ scaling of direct shot sampling to scaling close to $N_{\rm queries}^{-1}$. Our circuit-resource analysis shows that one step of the NVE propagator can be built with $\order{n^2}$ CX gates, where $n=n_x+n_p$ is the total number of position and momentum qubits. For the NVT propagator, the centered-difference Pauli-decomposition implementation of the Nos\'e--Hoover friction term scales as $\bigO(n_\xi n_p\,2^{n_p})$, where $n_p$ and $n_\xi$ are the numbers of momentum and thermostat qubits, respectively.
The proposed framework is a concrete step toward translating the principles of quantum algorithms into the transport-coefficient calculations required in practical molecular simulation.
\end{abstract}

\maketitle

\section{Introduction}\label{sec:intro}

Molecular dynamics (MD) is a powerful simulation tool for evaluating equilibrium structures, dynamical correlations, and transport coefficients by numerically integrating the equations of motion of atomic nuclei. Diffusion and viscosity coefficients are obtained from Green--Kubo relations as time integrals of equilibrium flux correlation functions~\cite{Tuckerman2010}. Such calculations are routine in materials design. In lithium-ion batteries, for example, MD is used to evaluate Li$^+$ diffusion coefficients in cathode materials, solid electrolytes, and solid-electrolyte interphases, which govern charge--discharge rates and ionic conductivity~\cite{Fallahzadeh2015,Baktash2020,Muralidharan2018}. However, computing transport coefficients, including diffusion coefficients, requires long-time averaging of correlation functions, and convergence of statistical errors requires either long trajectories or many independent samples.

Most work on simulating molecular systems with quantum computers has focused on electronic-structure calculations, the real-time evolution of quantum-chemical Hamiltonians, and the quantum dynamics of chemical reactions~\cite{Kassal2008,Whitfield2011,Su2021,Ollitrault2021}. Recently, frameworks based on the Koopman--von Neumann (KvN) representation have been proposed for treating classical Liouville dynamics on quantum computers, along with quantum algorithms for coupled quantum--classical dynamics involving classical nuclei and quantum electrons~\cite{Joseph2020,Simon2024,Huang2025Fullqubit}. These studies have mainly targeted properties of the equilibrium distribution, such as thermodynamic quantities and free-energy differences. By contrast, Green--Kubo transport coefficients require not only the equilibrium distribution but also the real-time correlation function of a state excited by a flux operator. In this respect, transport-coefficient evaluation is a readout problem distinct from equilibrium averages and free-energy calculations.

The requirement of physical-time evolution therefore guides the starting point of our formulation. Nos\'e's extended Hamiltonian provides a standard way to treat the canonical distribution as a Hamiltonian flow, and it has also been used in Liouvillian-type quantum algorithms for canonical ensembles and free-energy calculations~\cite{Nose1984,Simon2024,Huang2025Fullqubit}. However, the natural evolution parameter of the Nos\'e Hamiltonian is a scaled auxiliary time rather than physical time. For Green--Kubo integrals, however, the time variable entering the correlation function must be physical time. To avoid this mismatch, we start from the Nos\'e--Hoover equations, in which the time scaling has been absorbed~\cite{Hoover1985,Tuckerman2010}. Nos\'e--Hoover dynamics are formulated in physical time, but their phase-space flow is compressible. Consequently, a Hermitian KvN generator must be derived from the generalized Liouville equation rather than from the incompressible Hamiltonian Liouville operator.

The KvN representation expresses classical Liouville evolution as linear unitary evolution in Hilbert space~\cite{Koopman1931,vonNeumann1932}. This linearization allows phase-space distributions, flux operators, and time-correlation functions to be treated in the same Hilbert space. In this paper, we use this structure to map Green--Kubo transport coefficients in MD to a readout problem in quantum phase estimation (QPE).

The main results are threefold. First, we derive KvN generators for $\NVE$ dynamics and Nos\'e--Hoover-type $\NVT$ dynamics within the same framework and construct the corresponding one-step quantum circuits. Second, assuming that a canonical-state-preparation oracle provides the equilibrium state, we show that, when the flux-excited state is input to QPE, the bin-zero probability $P_0$ that the probability $P_0$ of measuring the QPE ancilla register in the all-zero state corresponds to a Bartlett-windowed Green--Kubo integral. Third, we numerically verify that maximum-likelihood amplitude estimation (MLAE) improves the statistical estimation of this $P_0$ from the $N_{\rm queries}^{-1/2}$ scaling of direct shot sampling to scaling close to $N_{\rm queries}^{-1}$~\cite{BrassardHoyer2002,Suzuki2020,Grinko2021}. Our resource analysis also shows that one step of the $\NVE$ propagator can be implemented with $\bigO(n^2)$ CX gates, where $n=n_x+n_p$ is the total number of position and momentum qubits. For the NVT propagator, the centered-difference Pauli-decomposition implementation of the Nos\'e--Hoover friction term scales as $\order{n_\xi n_p\,2^{n_p}}$, where $n_p$ and $n_\xi$ are the numbers of momentum and thermostat qubits, respectively.

\begin{figure}[t]
  \centering
  \includegraphics[width=\columnwidth]{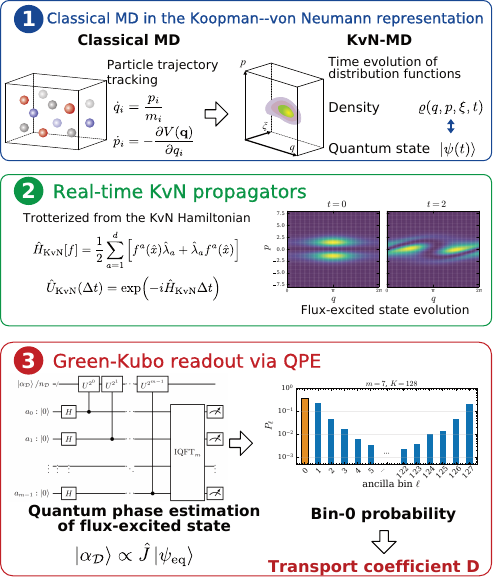}
  \caption{\textbf{Conceptual diagram of Green--Kubo transport-coefficient readout using KvN-MD and QPE.}
  The phase-space density of classical MD is represented as a KvN wave function, and $\NVE$ or Nos\'e--Hoover-type $\NVT$ dynamics are treated as a real-time KvN propagator. When the time evolution of the flux-excited state is input to QPE, a Bartlett-windowed Green--Kubo estimator is obtained from the bin-zero probability of the ancilla register.}
  \label{fig:pipeline}
\end{figure}

Figure~\ref{fig:pipeline} conceptually illustrates the readout workflow used in this paper. The real-time dynamics of classical MD are mapped to unitary evolution in KvN-MD, and the time evolution of the flux-excited state is input to QPE. As shown below, the bin-zero probability in QPE corresponds to a Bartlett-windowed Green--Kubo integral.

\section{Formulation of KvN-MD with a Nos\'e--Hoover Thermostat}\label{sec:kvn-formulation}

In this section, we map classical molecular dynamics to the Koopman--von Neumann (KvN) representation and describe both $\NVE$ and Nos\'e--Hoover-type $\NVT$ dynamics as unitary time evolutions on a Hilbert space. The KvN representation originates from the Hilbert-space formulation of classical mechanics by Koopman and von Neumann, and it has recently been revisited as a framework for treating classical nonlinear dynamics on quantum computers~\cite{Koopman1931,vonNeumann1932,Bondar2019,Joseph2020}.
We denote the physical phase-space variables by
$\{q_i,p_i\}_{i=1}^{N_f}$, where $N_f$ is the number of physical degrees of
freedom. Here, $q_i$ and $p_i$ are the $i$-th position and momentum. In the KvN Hilbert space, hats denote operators acting on phase-space
amplitudes. The operators $\hat q_i$, $\hat p_i$, and $\hat\xi$ act by
multiplication, while their conjugate derivative operators are defined as
\begin{equation}
  \hat\lambda_{q_i}
  =
  -i\frac{\partial}{\partial q_i},
  \qquad
  \hat\lambda_{p_i}
  =
  -i\frac{\partial}{\partial p_i}.
  \label{eq:kvn-derivative-operators}
\end{equation}
They satisfy
\begin{equation}
  [\hat q_i,\hat\lambda_{q_j}]
  =
  i\delta_{ij},
  \qquad
  [\hat p_i,\hat\lambda_{p_j}]
  =
  i\delta_{ij},
  \label{eq:kvn-canonical-commutators}
\end{equation}
with all other canonical commutators between different phase-space coordinates
vanishing. The variable $\xi$ is the Nos\'e--Hoover thermostat friction
variable.

Our aim is to construct a Hermitian KvN generator for compressible phase-space flows generated by thermostat dynamics. In particular, the Nos\'e--Hoover friction term $-\xi p_i$ gives rise to
\begin{equation}
  \hat H_3=-\hat\xi\sum_{i=1}^{N_f}\hat D_{p_i},
  \qquad
  \hat D_{p_i}=\frac{1}{2}\left(\hat p_i\hat\lambda_{p_i}+\hat\lambda_{p_i}\hat p_i\right).
  \label{eq:H3-preview}
\end{equation}
This dilation term is the main operator that distinguishes $\NVE$ from $\NVT$ in the circuit-resource analysis below.

\subsection{Generalized Liouville Equation and the KvN Representation}\label{sec:liouville-kvn}

Let $x=(x^1,\ldots,x^d)$ be a coordinate vector that specifies a point in the classical phase space. Here, $d$ is the dimension of the phase space used to describe the classical flow. For example, for physical $\NVE$ dynamics with $N_f$ degrees of freedom, one may take $x=(q_1,\ldots,q_{N_f},p_1,\ldots,p_{N_f})$ and hence $d=2N_f$. For Nos\'e--Hoover-type thermostat dynamics, $x$ may additionally include thermostat variables such as $\xi$, and $d$ is enlarged accordingly. 
We write the deterministic equations of motion as 
\begin{equation} \dot x^a = f^a(x), \qquad a=1,\ldots,d. \label{eq:classical-flow} \end{equation} 
Here, $f(x)=(f^1(x),\ldots,f^d(x))$ is the phase-space velocity vector field, and $f^a(x)$ denotes its component along the coordinate $x^a$. Throughout this section, we use the shorthand $\partial_a=\partial/\partial x^a$. Let $\varrho(x,t)$ denote the probability density at the phase-space point $x$ and time $t$, defined with respect to the flat phase-space volume element $\mathrm{d}^d x=\mathrm{d}x^1\cdots \mathrm{d}x^d$. Conservation of probability is expressed by the continuity equation 
\begin{equation} \frac{\partial \varrho(x,t)}{\partial t} + \sum_{a=1}^{d} \frac{\partial}{\partial x^a} \left[ f^a(x)\varrho(x,t) \right] = 0. \label{eq:liouville-flat} 
\end{equation}

This form of the Liouville equation applies also to compressible, non-Hamiltonian flows. Expanding the divergence term gives
\begin{align}
\frac{\partial \varrho(x,t)}{\partial t} &= -\sum_{a=1}^{d} f^a(x)\partial_a\varrho(x,t) - \kappa(x)\varrho(x,t), 
\nonumber \\ 
\kappa(x) &= \sum_{a=1}^{d}\partial_a f^a(x). 
\label{eq:liouville-compressibility} 
\end{align}
Here, $\kappa(x)$ is the phase-space compressibility, namely the divergence of the flow field $f(x)$. For canonical Hamiltonian dynamics, $\kappa(x)=0$, whereas thermostat dynamics can give $\kappa(x)\neq 0$.
The compressibility term is therefore essential. Omitting it would break probability conservation at the density level and, in the KvN representation, would correspond to using a non-Hermitian generator~\cite{Tuckerman2010}.

More generally, one may introduce a weighted phase-space measure $\mu(x)\mathrm{d}^d x$, with $\mu(x)>0$, and regard $\rho(x,t)$ as a density with respect to that measure. The density with respect to the flat measure is then $\varrho=\mu\rho$, and the generalized Liouville equation can be written as
\begin{equation}
  \frac{\partial}{\partial t}\left[\mu(x)\rho(x,t)\right]
  +\sum_{a=1}^{d}\frac{\partial}{\partial x^a}
  \left[\mu(x)\rho(x,t)f^a(x)\right]=0.
  \label{eq:generalized-liouville}
\end{equation}
Equivalently, in terms of $\rho$, this equation becomes
\begin{equation}
  \begin{aligned}
    \frac{\partial \rho(x,t)}{\partial t}
    &=
    -\sum_{a=1}^{d}
    f^a(x)\partial_a\rho(x,t)
    -
    \rho(x,t)\,\kappa_\mu(x),
    \\
    \kappa_\mu(x)
    &=
    \frac{1}{\mu(x)}
    \sum_{a=1}^{d}
    \partial_a
    \left[
      \mu(x) f^a(x)
    \right].
  \end{aligned}
  \label{eq:generalized-liouville-expanded}
\end{equation}
In particular, a weight $\mu$ satisfying $\kappa_\mu=0$ defines an invariant measure of the flow, and the density $\rho$ with respect to that measure obeys a pure advection equation. In the quantum registers used in this work, however, amplitudes on the discrete grid are normalized so as to represent the square root of the physical density $\varrho$ with respect to the flat measure. We therefore use, in what follows, the Hermitian generator corresponding to Eq.~\eqref{eq:liouville-flat}.

In the KvN representation, the probability density is written as the squared modulus of a wave function,
\begin{equation}
  \varrho(x,t)=|\psi(x,t)|^2.
  \label{eq:kvn-density}
\end{equation}
Defining the operator conjugate to $x^a$ by
\begin{equation}
  \hat\lambda_a
  =
  -\mathrm{i}\frac{\partial}{\partial x^a},
  \qquad
  [\hat x^a,\hat\lambda_b]
  =
  \mathrm{i}\delta_{ab}.
  \label{eq:lambda-commutator}
\end{equation}
the Hermitian KvN generator that reproduces Eq.~\eqref{eq:liouville-flat} is~\cite{Bondar2019,Joseph2020}
\begin{equation}
\begin{aligned}
  \hat H_{\rm KvN}[f]
  &=
  \frac{1}{2}\sum_{a=1}^{d}
  \left[
    f^a(\hat x)\hat\lambda_a
    +
    \hat\lambda_a f^a(\hat x)
  \right] \\
  &=
  \sum_{a=1}^{d} f^a(\hat x)\hat\lambda_a
  -
  \frac{\mathrm{i}}{2}
  \sum_{a=1}^{d}\partial_a f^a(\hat x).
\end{aligned}
\label{eq:general-kvn-generator}
\end{equation}
Indeed, the Schr\"odinger-type equation generated by this operator,
\begin{equation}
  \mathrm{i}\frac{\partial}{\partial t}\ket{\psi(t)}
  =
  \hat H_{\rm KvN}[f]\ket{\psi(t)}.
  \label{eq:kvn-schrodinger}
\end{equation}
In the coordinate representation, the KvN wavefunction is defined as
$\psi(x,t)=\braket{x}{\psi(t)}$. Equation~\eqref{eq:kvn-schrodinger}
then implies
\begin{equation}
  \frac{\partial \psi(x,t)}{\partial t}
  =
  -\sum_{a=1}^{d}
  f^a(x)\partial_a\psi(x,t)
  -
  \frac{1}{2}
  \left[
    \sum_{a=1}^{d}\partial_a f^a(x)
  \right]
  \psi(x,t).
  \label{eq:kvn-wavefunction-continuity}
\end{equation}
Equivalently, using the phase-space compressibility
$\kappa(x)=\sum_{a=1}^{d}\partial_a f^a(x)$, this equation can be written as
\begin{equation}
  \frac{\partial \psi(x,t)}{\partial t}
  =
  -\sum_{a=1}^{d}
  f^a(x)\partial_a\psi(x,t)
  -
  \frac{1}{2}\kappa(x)\psi(x,t).
  \label{eq:kvn-wavefunction-compressibility}
\end{equation}
Therefore, the probability density
$\varrho(x,t)=|\psi(x,t)|^2$ obeys Eq.~\eqref{eq:liouville-flat}. When
$\kappa(x)=0$, as in canonical Hamiltonian flows, Eq.~\eqref{eq:general-kvn-generator}
reduces to the usual incompressible-flow Liouville operator
$\sum_{a=1}^{d} f^a(\hat x)\hat\lambda_a$, where
$\hat\lambda_a=-\mathrm{i}\partial_a$ in the $x$ representation.

For the physical phase-space variables, we write
$x=(q,p)=(q_1,\ldots,q_{N_f},p_1,\ldots,p_{N_f})$, where $q_i$ and $p_i$
are the $i$th coordinate and its conjugate momentum. Thus, $x$ is the
phase-space coordinate vector, while $q$ and $p$ denote its coordinate and
momentum components, respectively.

A key feature of this construction is that the position operators $\hat q_i$
and momentum operators $\hat p_i$ are not the noncommuting canonical operators
of quantum mechanics. Rather, they are mutually commuting classical
phase-space coordinates represented as multiplication operators on the KvN
Hilbert space. Their conjugate derivative operators are
\begin{equation}
  [\hat q_i,\hat p_j]=0,
  \qquad
  [\hat q_i,\hat\lambda_{q_j}]=\mathrm{i}\delta_{ij},
  \qquad
  [\hat p_i,\hat\lambda_{p_j}]=\mathrm{i}\delta_{ij}.
  \label{eq:kvn-commutators}
\end{equation}
A classical observable $A(q,p)$ acts as the multiplication operator
$\hat A=A(\hat q,\hat p)$ on the KvN Hilbert space. In the $(q,p)$
representation, the KvN wavefunction is
$\psi(q,p,t)=\braket{q,p}{\psi(t)}$, and the associated probability density is
$\varrho(q,p,t)=|\psi(q,p,t)|^2$. The corresponding classical average at time
$t$ is therefore written as
\begin{equation} 
\langle A\rangle_{\varrho(t)} = \int A(q,p)\varrho(q,p,t)\, \mathrm{d}^{N_f}q\,\mathrm{d}^{N_f}p = \bra{\psi(t)}\hat A\ket{\psi(t)}. \label{eq:observable-average} 
\end{equation}
This correspondence between the classical probability density
$\varrho(q,p,t)$ and the Hilbert-space norm of $\ket{\psi(t)}$ is what later
allows Green--Kubo correlation functions to be read as inner products in a
quantum circuit.

\subsection{NVE KvN Generator}\label{sec:nve-kvn}

We first consider $\NVE$ dynamics for the physical phase-space variables
$\mathbf{q}=(q_1,\ldots,q_{N_f})$ and
$\mathbf{p}=(p_1,\ldots,p_{N_f})$, where $N_f$ is the number of classical
degrees of freedom. The classical Hamiltonian is
\begin{equation}
  H_{\rm cl}(\mathbf{q},\mathbf{p})
  =
  \sum_{i=1}^{N_f}
  \frac{p_i^2}{2m_i}
  +
  V(\mathbf{q}).
  \label{eq:classical-hamiltonian}
\end{equation}
The system evolves on the potential-energy surface $V(\mathbf{q})$. The force
acting on the $i$th degree of freedom is defined by
\begin{equation}
  F_i(\mathbf{q})
  =
  -\frac{\partial V(\mathbf{q})}{\partial q_i}.
  \label{eq:nve-force}
\end{equation}
The equations of motion are therefore
\begin{equation}
  \dot q_i
  =
  \frac{p_i}{m_i},
  \qquad
  \dot p_i
  =
  F_i(\mathbf{q}),
  \qquad
  i=1,\ldots,N_f.
  \label{eq:nve-eom}
\end{equation}
This Hamiltonian flow is divergence-free in the phase-space coordinates
$(\mathbf{q},\mathbf{p})$. Hence, the KvN generator contains no
compressibility term and is given by
\begin{equation}
  \hat H_{\rm KvN}^{\rm NVE}
  =
  \sum_{i=1}^{N_f}
  \frac{\hat p_i}{m_i}\hat\lambda_{q_i}
  +
  \sum_{i=1}^{N_f}
  F_i(\hat{\mathbf{q}})\hat\lambda_{p_i}
  \equiv
  \hat H_1+\hat H_2.
  \label{eq:nve-kvn-generator}
\end{equation}

\subsection{KvN Generator for the Nos\'e--Hoover Thermostat}\label{sec:nose-hoover-kvn}

To describe the canonical ensemble, we introduce the Nos\'e--Hoover thermostat
variable $\xi$ in addition to the physical phase-space variables
$\mathbf{q}=(q_1,\ldots,q_{N_f})$ and
$\mathbf{p}=(p_1,\ldots,p_{N_f})$~\cite{Nose1984,Hoover1985,Tuckerman2010}.
Here, $q_i$ and $p_i$ are the position and momentum of the $i$th
degree of freedom, respectively. We denote the extended phase-space coordinate
by $\mathbf{z}=(\mathbf{q},\mathbf{p},\xi)$.

The single-stage Nos\'e--Hoover equations are
\begin{equation}
  \begin{aligned}
    \dot q_i
    &=
    \frac{p_i}{m_i},
    \\
    \dot p_i
    &=
    F_i(\mathbf{q})-\xi p_i,
    \\
    \dot\xi
    &=
    \frac{1}{Q}
    \left(
      \sum_{i=1}^{N_f}\frac{p_i^2}{m_i}
      -
      N_f T_0
    \right).
  \end{aligned}
  \label{eq:nh-eom}
\end{equation}
Here, $Q$ is the
thermostat mass, $T_0$ is the target temperature, and we use units with
$k_{\rm B}=1$. Because of the friction term $-\xi p_i$, the momenta are
damped for $\xi>0$ and amplified for $\xi<0$. Equation~\eqref{eq:nh-eom}
defines a deterministic flow on the extended phase space,
\begin{equation}
  \dot{\mathbf{z}}
  =
  \mathbf{f}_{\rm NH}(\mathbf{z}),
  \label{eq:nh-flow}
\end{equation}
where $\mathbf{f}_{\rm NH}(\mathbf{z})$ is the Nos\'e--Hoover vector fieldd
whose components are given by the right-hand sides of Eq.~\eqref{eq:nh-eom}.

The phase-space compressibility of the Nos\'e--Hoover flow is the divergence
of $\mathbf{f}_{\rm NH}(\mathbf{z})$ in the extended variables
$\mathbf{z}=(\mathbf{q},\mathbf{p},\xi)$:
\begin{equation}
  \begin{aligned}
    \kappa_{\rm NH}(\mathbf{z})
    &=
    -N_f\xi.
  \end{aligned}
  \label{eq:nh-compressibility}
\end{equation}
For the extended canonical density
\begin{equation}
  \begin{aligned}
    \varrho_{\rm ext}(\mathbf{q},\mathbf{p},\xi)
    &\propto
    \exp
    \left[
      -\beta
      \left(
        H_{\rm cl}(\mathbf{q},\mathbf{p})
        +
        \frac{Q\xi^2}{2}
      \right)
    \right],
    \\
    \beta
    &=
    T_0^{-1},
  \end{aligned}
  \label{eq:extended-canonical-density}
\end{equation}
a direct calculation gives
\begin{equation}
  \mathbf{f}_{\rm NH}(\mathbf{z})
  \cdot
  \nabla_{\mathbf z}
  \log \varrho_{\rm ext}(\mathbf{q},\mathbf{p},\xi)
  =
  N_f\xi
  =
  -\kappa_{\rm NH}(\mathbf{z}).
  \label{eq:nh-stationary-check}
\end{equation}
It follows from Eq.~\eqref{eq:liouville-compressibility} that
\begin{equation}
  \nabla_{\mathbf z}
  \cdot
  \left[
    \mathbf{f}_{\rm NH}(\mathbf{z})
    \varrho_{\rm ext}(\mathbf{q},\mathbf{p},\xi)
  \right]
  =
  0.
  \label{eq:nh-invariant-density}
\end{equation}
Thus, Eq.~\eqref{eq:extended-canonical-density} is an invariant density of
the Nos\'e--Hoover flow. After integrating out $\xi$, the marginal distribution
on $(\mathbf{q},\mathbf{p})$ is the usual canonical distribution proportional
to $\exp[-\beta H_{\rm cl}(\mathbf{q},\mathbf{p})]$.

The existence of an invariant density, however, does not imply that the density
can be sampled efficiently from an arbitrary initial state. The ergodicity of a
single-stage Nos\'e--Hoover trajectory can fail in low-dimensional or strongly
integrable systems, and Nos\'e--Hoover chains are often used for practical
canonical sampling~\cite{MartynaKlein1992,Tuckerman2010}. In this work, we use
the single-stage Nos\'e--Hoover equations as the reference case in order to
clarify the structure of the KvN generator and its quantum-circuit
implementation for thermostatted classical flows. In an extension to a
multistage chain, dilation-type terms of the same kind arise from the couplings
between thermostat variables. Thus, the essential circuit-level difficulty is
already represented by the $\hat H_3$ term derived below.

Because Eq.~\eqref{eq:nh-eom} defines a compressible flow on the extended
phase space $\mathbf{z}=(\mathbf{q},\mathbf{p},\xi)$, the unsymmetrized
Liouville operator used for divergence-free $\NVE$ dynamics is no longer
Hermitian. We therefore apply the symmetrized KvN construction in
Eq.~\eqref{eq:general-kvn-generator} to each component of the Nos\'e--Hoover
vector field $\mathbf{f}_{\rm NH}(\mathbf{z})$. The contributions from the
$\mathbf{q}$-component advection, the conservative force-induced
$\mathbf{p}$-component advection, and the $\xi$-component advection are,
respectively,
\begin{equation}
  \hat H_1
  =
  \sum_{i=1}^{N_f}
  \frac{\hat p_i}{m_i}\hat\lambda_{q_i}.
  \label{eq:H1-def}
\end{equation}
\begin{equation}
  \hat H_2
  =
  \sum_{i=1}^{N_f}
  F_i(\hat{\mathbf{q}})\hat\lambda_{p_i}.
  \label{eq:H2-def}
\end{equation}
\begin{equation}
  \hat H_4
  =
  \frac{1}{Q}
  \left(
    \sum_{i=1}^{N_f}
    \frac{\hat p_i^2}{m_i}
    -
    N_f T_0
  \right)
  \hat\lambda_\xi.
  \label{eq:H4-def}
\end{equation}
Because each coefficient in
Eqs.~\eqref{eq:H1-def}--\eqref{eq:H4-def} is independent of its corresponding
differentiation variable, the symmetrization generates no additional terms for
these three contributions.

The nontrivial contribution comes from the friction part
$-\xi p_i$ of the $p_i$ equation of motion. Applying
Eq.~\eqref{eq:general-kvn-generator} to this component gives
\begin{equation}
  \begin{aligned}
    \frac{1}{2}
    \left[
      (-\hat\xi\hat p_i)\hat\lambda_{p_i}
      +
      \hat\lambda_{p_i}(-\hat\xi\hat p_i)
    \right]
    &=
    -\hat\xi\,
    \frac{1}{2}
    \left(
      \hat p_i\hat\lambda_{p_i}
      +
      \hat\lambda_{p_i}\hat p_i
    \right).
  \end{aligned}
  \label{eq:friction-symmetrization}
\end{equation}
Thus, defining the Hermitian dilation generator in the $p_i$ direction as
\begin{equation}
  \hat D_{p_i}
  \equiv
  \frac{1}{2}
  \left(
    \hat p_i\hat\lambda_{p_i}
    +
    \hat\lambda_{p_i}\hat p_i
  \right),
  \label{eq:Dpi-def}
\end{equation}
the thermostat-specific contribution is
\begin{equation}
  \hat H_3
  =
  -\hat\xi
  \sum_{i=1}^{N_f}
  \hat D_{p_i}.
  \label{eq:H3-multidim}
\end{equation}
Combining these terms, the $\NVT$ KvN generator decomposes as
\begin{equation}
  \hat H_{\rm KvN}^{\rm NVT}
  =
  \hat H_1
  +
  \hat H_2
  +
  \hat H_3
  +
  \hat H_4.
  \label{eq:nvt-kvn-generator}
\end{equation}

\subsection{Meaning of the Dilation Operator and NVT Time Evolution}\label{sec:dilation-and-nvt-product}

The operator $\hat D_p$ is the generator of dilations in momentum space. In the continuous representation,
\begin{equation}
  \hat D_p
  =\frac{1}{2}\left(p(-\mathrm{i}\partial_p)+(-\mathrm{i}\partial_p)p\right)
  =-\mathrm{i}\left(p\partial_p+\frac{1}{2}\right).
  \label{eq:Dp-continuum}
\end{equation}
It follows that
\begin{equation}
  e^{-\mathrm{i} a\hat D_p}\psi(p)
  =e^{-a/2}\psi(e^{-a}p),
  \label{eq:dilation-action}
\end{equation}
so the wave-function amplitude is rescaled along the momentum axis while the norm is preserved. For the partial evolution generated by $\hat H_3=-\xi\hat D_p$, one has $a=-\xi\Delta t$. Hence a peak initially located at momentum $p$ is shifted to
\[
  p(t+\Delta t)=e^{-\xi\Delta t}p(t),
\]
in agreement with the classical solution of $\dot p=-\xi p$. This corresponds to cooling for $\xi>0$ and heating for $\xi<0$.

Starting from the generalized Liouville equation, we have thus obtained Hermitian KvN generators for both $\NVE$ and Nos\'e--Hoover-type $\NVT$ dynamics. In the quantum circuits below, the corresponding one-step propagators serve as the basic building blocks.


\section{Quantum Circuit Implementation}\label{sec:circuit-implementation}

In this section, we translate the KvN generators derived in the previous section into quantum circuit blocks for their one-step time evolutions.
To keep the notation compact for many-particle systems, we collect all position and momentum degrees of freedom into the bundled registers
\begin{equation}
  |\mathbf q\rangle=|q_1,\ldots,q_{N_f}\rangle,
  \qquad
  |\mathbf p\rangle=|p_1,\ldots,p_{N_f}\rangle .
  \label{eq:multi-particle-registers}
\end{equation}
We denote the corresponding total numbers of qubits by $n_q$ and $n_p$.
For Nos\'e--Hoover-type $\NVT$ dynamics, we also introduce the thermostat register $|\xi\rangle$, with $n_\xi$ qubits.
The operators $\mathcal F_{\mathbf q}$ and $\mathcal F_{\mathbf p}$ denote tensor products of QFTs acting on all $q_i$ and $p_i$ registers, respectively.
The operators $\hat k_{q_i}$ and $\hat k_{p_i}$ are the corresponding discrete conjugate wavenumber operators in the Fourier-transformed bases.
When the thermostat register is Fourier transformed, we use the analogous notation $\mathcal F_\xi$ and $\hat k_\xi$.

\subsection{One-Step Propagator}
\label{sec:multi-particle-step-circuit}

For each partial generator $\hat H_j$, we define the corresponding
short-time propagator as
\begin{equation}
  U_j(\delta t)
  =
  \exp[-\ii\delta t\hat H_j].
  \label{eq:Uj-definition-circuit}
\end{equation}
For the time discretization, we use the standard second-order symmetric
Suzuki--Trotter product formula of quantum simulation~\cite{Trotter1959,Suzuki1976,Suzuki1990,Yoshida1990,Kivlichan2019}.

For $\NVE$ dynamics, the KvN generator is decomposed as
$\hat H_{\rm KvN}^{\NVE}=\hat H_1+\hat H_2$, where
\begin{equation}
  \begin{aligned}
    \hat H_1
    &=
    \sum_{i=1}^{N_f}
    \frac{\hat p_i}{m_i}\hat\lambda_{q_i},
    \\
    \hat H_2
    &=
    \sum_{i=1}^{N_f}
    F_i(\hat{\mathbf q})\hat\lambda_{p_i}.
  \end{aligned}
  \label{eq:nve-partial-generators-circuit}
\end{equation}
The terms within each sum mutually commute. Therefore, the many-degree-of-freedom
propagator can be written in the same drift--kick--drift form as in the
single-degree-of-freedom case, with all position degrees of freedom updated
collectively by a half step, all momentum degrees of freedom updated
collectively by a full step, and the position degrees of freedom updated again
by a half step:
\begin{equation}
  S_2^{\NVE}(\Delta t)
  =
  U_1(\Delta t/2)
  U_2(\Delta t)
  U_1(\Delta t/2).
  \label{eq:nve-circuit-step}
\end{equation}
By the Baker--Campbell--Hausdorff expansion, this palindromic product formula
satisfies
\begin{equation}
  S_2^{\NVE}(\Delta t)
  =
  \exp
  \left[
    -\ii
    \left(
      \hat H_1+\hat H_2
    \right)
    \Delta t
    +
    \bigO(\Delta t^3)
  \right].
  \label{eq:nve-circuit-step-error}
\end{equation}
Equivalently, the local error of one step is $\bigO(\Delta t^3)$, and the
global error up to a fixed physical time is $\bigO(\Delta t^2)$. This
time-reversal symmetry is useful for the long-time correlation functions
considered in this work.

For Nos\'e--Hoover-type $\NVT$ dynamics, the KvN generator is decomposed as
$\hat H_{\rm KvN}^{\NVT}=\hat H_1+\hat H_2+\hat H_3+\hat H_4$, where
$\hat H_3$ contains the momentum-dilation generator and $\hat H_4$ generates
the thermostat-variable advection:
\begin{equation}
  \begin{aligned}
    \hat H_3
    &=
    -\hat\xi
    \sum_{i=1}^{N_f}
    \hat D_{p_i},
    \\
    \hat H_4
    &=
    \frac{1}{Q}
    \left(
      \sum_{i=1}^{N_f}
      \frac{\hat p_i^2}{m_i}
      -
      N_fT_0
    \right)
    \hat\lambda_{\xi}.
  \end{aligned}
  \label{eq:nvt-additional-generators-circuit}
\end{equation}
Using the same second-order symmetric Suzuki--Trotter construction, we take
\begin{equation}
  \begin{aligned}
    S_2^{\NVT}(\Delta t)
    &=
    U_1(\Delta t/2)
    U_2(\Delta t/2)
    U_3(\Delta t/2)
    U_4(\Delta t)
    \\
    &\quad\times
    U_3(\Delta t/2)
    U_2(\Delta t/2)
    U_1(\Delta t/2).
  \end{aligned}
  \label{eq:nvt-circuit-step}
\end{equation}
This palindromic product formula satisfies
\begin{equation}
  S_2^{\NVT}(\Delta t)
  =
  \exp
  \left[
    -\ii
    \left(
      \hat H_1+\hat H_2+\hat H_3+\hat H_4
    \right)
    \Delta t
    +
    \bigO(\Delta t^3)
  \right].
  \label{eq:nvt-circuit-step-error}
\end{equation}
Thus, the local error of one $\NVT$ step is also $\bigO(\Delta t^3)$, and
the global error up to a fixed physical time is $\bigO(\Delta t^2)$. The term
$\hat H_3$ is the central additional component of the real-time $\NVT$
propagator relative to the $\NVE$ case, because it contains the
momentum-dilation generator induced by the Nos\'e--Hoover friction. Its
quantum-circuit implementation is discussed in the next section.

\begin{figure*}[t]
  \centering

  \begin{minipage}[t]{0.55\textwidth}
    \vspace{0pt}
    \centering

    \begin{adjustbox}{max width=\linewidth,center}
    \begin{quantikz}[font=\small, column sep=0.13cm, row sep=0.40cm, line width=0.55pt]
      \lstick{$\ket{\mathbf q}/n_q$}
        & \qw \push{\scriptstyle /}
        & \gate{\mathcal{F}_{\mathbf q}}
        & \gate[wires=2]{\Phi_q(\Delta t/2)}
        & \gate{\mathcal{F}_{\mathbf q}^{\dagger}}
        & \qw
        & \gate[wires=2]{\Phi_p(\Delta t)}
        & \qw
        & \gate{\mathcal{F}_{\mathbf q}}
        & \gate[wires=2]{\Phi_q(\Delta t/2)}
        & \gate{\mathcal{F}_{\mathbf q}^{\dagger}}
        & \qw \\
      \lstick{$\ket{\mathbf p}/n_p$}
        & \qw \push{\scriptstyle /}
        & \qw
        & \ghost{\Phi_q(\Delta t/2)}
        & \qw
        & \gate{\mathcal{F}_{\mathbf p}}
        & \ghost{\Phi_p(\Delta t)}
        & \gate{\mathcal{F}_{\mathbf p}^{\dagger}}
        & \qw
        & \ghost{\Phi_q(\Delta t/2)}
        & \qw
        & \qw
    \end{quantikz}
    \end{adjustbox}

    \vspace{0.32em}
    {\small
    \textbf{(a)} NVE one-step circuit:
    $S_2^{\mathrm{NVE}}(\Delta t)
    =U_1(\Delta t/2)U_2(\Delta t)U_1(\Delta t/2)$.
    }

    \vspace{2.0em}

    \begin{adjustbox}{max width=\linewidth,center}
    \begin{quantikz}[font=\small, column sep=0.11cm, row sep=0.40cm, line width=0.55pt]
      \lstick{$\ket{\mathbf q}/n_q$}
        & \qw \push{\scriptstyle /}
        & \gate[wires=2]{U_1(\Delta t/2)}
        & \gate[wires=2]{U_2(\Delta t/2)}
        & \qw
        & \qw
        & \qw
        & \qw
        & \qw
        & \gate[wires=2]{U_2(\Delta t/2)}
        & \gate[wires=2]{U_1(\Delta t/2)}
        & \qw \\
      \lstick{$\ket{\mathbf p}/n_p$}
        & \qw \push{\scriptstyle /}
        & \ghost{U_1(\Delta t/2)}
        & \ghost{U_2(\Delta t/2)}
        & \gate[wires=2]{U_3(\Delta t/2)}
        & \qw
        & \gate[wires=2]{\Phi_4(\Delta t)}
        & \qw
        & \gate[wires=2]{U_3(\Delta t/2)}
        & \ghost{U_2(\Delta t/2)}
        & \ghost{U_1(\Delta t/2)}
        & \qw \\
      \lstick{$\ket{\xi}/n_{\xi}$}
        & \qw \push{\scriptstyle /}
        & \qw
        & \qw
        & \ghost{U_3(\Delta t/2)}
        & \gate{\mathcal{F}_{\xi}}
        & \ghost{\Phi_4(\Delta t)}
        & \gate{\mathcal{F}_{\xi}^{\dagger}}
        & \ghost{U_3(\Delta t/2)}
        & \qw
        & \qw
        & \qw
    \end{quantikz}
    \end{adjustbox}

    \vspace{0.32em}
    {\small
    \textbf{(b)} NVT one-step circuit:
    $S_2^{\mathrm{NVT}}(\Delta t)=U_1U_2U_3U_4U_3U_2U_1$.
    }

  \end{minipage}
  \hfill
  \begin{minipage}[t]{0.41\textwidth}
    \vspace{0pt}
    \centering
    \vspace{1.2em}

    \begin{adjustbox}{max width=\linewidth,center}
    \begin{quantikz}[font=\small, column sep=0.10cm, row sep=0.50cm, line width=0.55pt]
      \lstick{$\ket{\alpha_{\mathcal D}}/n_{\mathcal D}$}
        & \qw \push{\scriptstyle /}
        & \qw
        & \gate{U^{2^0}}
        & \gate{U^{2^1}}
        & \cdots
        & \gate{U^{2^{m-1}}}
        & \qw
        & \qw \\
      \lstick{$a_0:\ket{0}$}
        & \qw
        & \gate{H}
        & \ctrl{-1}
        & \qw
        & \cdots
        & \qw
        & \gate[wires=4]{\mathrm{IQFT}_m}
        & \meter{} \\
      \lstick{$a_1:\ket{0}$}
        & \qw
        & \gate{H}
        & \qw
        & \ctrl{-2}
        & \cdots
        & \qw
        & \ghost{\mathrm{IQFT}_m}
        & \meter{} \\
      \lstick{$\vdots$}
        & \vdots
        & \vdots
        & \qw
        & \qw
        & \ddots
        & \qw
        & \ghost{\mathrm{IQFT}_m}
        & \vdots \\
      \lstick{$a_{m-1}:\ket{0}$}
        & \qw
        & \gate{H}
        & \qw
        & \qw
        & \cdots
        & \ctrl{-4}
        & \ghost{\mathrm{IQFT}_m}
        & \meter{}
    \end{quantikz}
    \end{adjustbox}

    \vspace{0.36em}
    {\small \textbf{(c)} QPE readout circuit.}

  \end{minipage}

  \vspace{0.65em}

  \caption{\textbf{Quantum circuits for KvN propagators and QPE readout.}
  (a) Multi-particle $\NVE$ one-step propagator in drift--kick--drift form.
  The bundled registers $\ket{\mathbf q}$ and $\ket{\mathbf p}$ contain all
  classical degrees of freedom, and the slash denotes a multi-qubit register.
  The QFTs $\mathcal F_{\mathbf q}$ and $\mathcal F_{\mathbf p}$ are tensor
  products over the corresponding coordinate registers. The diagonal phases are
  $\Phi_q(\delta t)=\exp[-\ii\delta t\sum_i \hat p_i\hat k_{q_i}/m_i]$ and
  $\Phi_p(\delta t)=\exp[-\ii\delta t\sum_i F_i(\hat{\mathbf q})\hat k_{p_i}]$.
  (b) Nos\'e--Hoover $\NVT$ one-step propagator. The additional dilation block is
  $U_3(\delta t)=\exp[-\ii\delta t\hat H_3]$, with
  $\hat H_3=-\hat\xi\sum_i\hat D_{p_i}$ and
  $\hat D_{p_i}=(\hat p_i\hat\lambda_{p_i}+\hat\lambda_{p_i}\hat p_i)/2$.
  The thermostat block is implemented as
  $\mathcal F_\xi^\dagger\Phi_4(\delta t)\mathcal F_\xi$, where
  $\Phi_4(\delta t)=
  \exp\{-\ii\delta t[(\sum_i\hat p_i^2/m_i)-N_fT_0]\hat k_\xi/Q\}$.
  (c) QPE readout circuit. The controlled powers of
  $U=S_2^{\mathcal D}(\Delta t)$ act on the flux-excited state
  $\ket{\alpha_{\mathcal D}}\propto \hat J\ket{\psi_{{\rm eq},\mathcal D}}$.
  Here $n_{\mathcal D}=n_q+n_p$ for $\mathcal D=\NVE$ and
  $n_{\mathcal D}=n_q+n_p+n_\xi$ for $\mathcal D=\NVT$. The bin-zero
  probability gives the Bartlett-windowed Green--Kubo integral derived in
  Sec.~\ref{sec:qpe-readout}.}
  \label{fig:circuits}
\end{figure*}
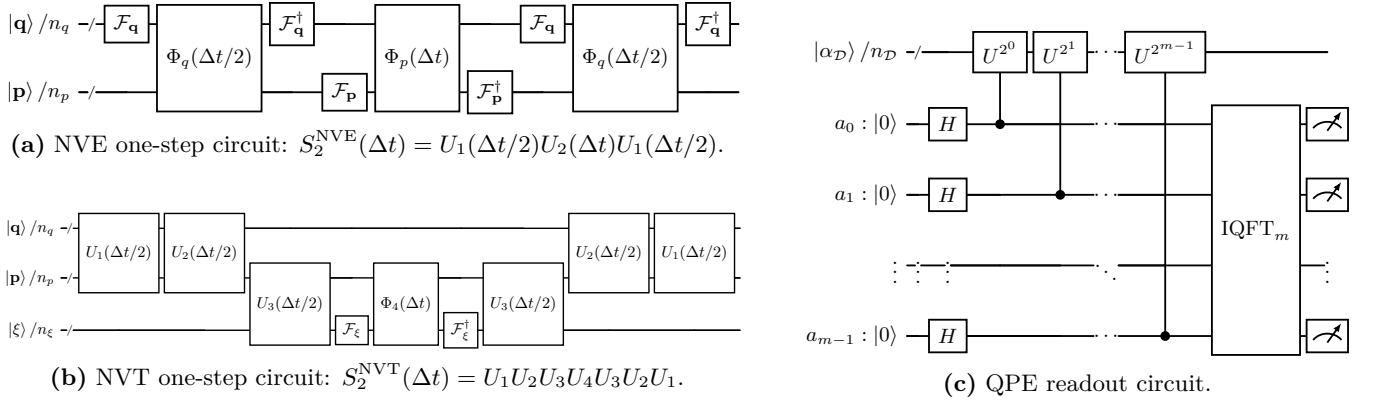

Figure~\ref{fig:circuits} shows these one-step propagators and the QPE readout
circuit.

Figure~\ref{fig:circuits}(a) uses diagonal unitaries that act collectively on
the bundled coordinate and momentum registers. For the drift generator
$\hat H_1$, the QFT is applied to the coordinate registers
$\mathbf{q}=(q_1,\ldots,q_{N_f})$, so that
$\hat\lambda_{q_i}$ is represented by the diagonal Fourier-conjugate operator
$\hat k_{q_i}$. The corresponding diagonal phase is
\begin{equation}
  \Phi_q(\delta t)
  =
  \exp
  \left[
    -\ii\delta t
    \sum_{i=1}^{N_f}
    \frac{\hat p_i}{m_i}
    \hat k_{q_i}
  \right].
  \label{eq:Phi-q-definition}
\end{equation}
Note that $\hat k_{q_i}$ is the Fourier-conjugate operator associated with
the $q_i$ register, and should not be confused with the physical momentum
coordinate $\hat p_i$.

For the force generator $\hat H_2$, the QFT is applied to the momentum
registers $\mathbf{p}=(p_1,\ldots,p_{N_f})$, so that
$\hat\lambda_{p_i}$ is represented by the diagonal Fourier-conjugate operator
$\hat k_{p_i}$. The corresponding diagonal phase is
\begin{equation}
  \Phi_p(\delta t)
  =
  \exp
  \left[
    -\ii\delta t
    \sum_{i=1}^{N_f}
    F_i(\hat{\mathbf q})
    \hat k_{p_i}
  \right].
  \label{eq:Phi-p-definition}
\end{equation}
The force component $F_i(\hat{\mathbf q})$ may depend on all coordinate
registers and is therefore implemented as a diagonal function of the bundled
coordinate register $\hat{\mathbf q}$.

For the thermostat-advection generator $\hat H_4$ in the $\NVT$ circuit, the
QFT is applied to the thermostat register $\xi$, so that
$\hat\lambda_\xi$ is represented by the diagonal Fourier-conjugate operator
$\hat k_\xi$. The corresponding diagonal phase is
\begin{equation}
  \Phi_4(\delta t)
  =
  \exp
  \left[
    -\ii\delta t
    \frac{
      \sum_{i=1}^{N_f}\hat p_i^2/m_i
      -
      N_fT_0
    }{Q}
    \hat k_\xi
  \right].
  \label{eq:Phi4-definition}
\end{equation}
Thus, $U_1$, $U_2$, and $U_4$ have the QFT--diagonal-phase--inverse-QFT
structure used in split-operator quantum simulation~\cite{Kassal2008,Welch2014}.
By contrast, $U_3$ contains the momentum-dilation generator
$\hat D_{p_i}$ and is not diagonalized by a simple QFT on a single register.
In this work, we implement $U_3$ using the centered-difference
Pauli-decomposition method described in
Appendix~\ref{app:central-difference-pauli}.

\subsection{State Preparation and the Flux-Excited State}
\label{sec:state-preparation-oracle}

The QPE circuit is initialized with a flux-excited KvN state. For
$\mathcal D\in\{\NVE,\NVT\}$, we define
\begin{equation}
  \begin{aligned}
    \mathbf z_{\mathcal D}
    &=
    \begin{cases}
      (\mathbf q,\mathbf p), & \mathcal D=\NVE, \\
      (\mathbf q,\mathbf p,\xi), & \mathcal D=\NVT,
    \end{cases}
    \\
    |\mathbf z_{\mathcal D}\rangle
    &=
    \begin{cases}
      |\mathbf q,\mathbf p\rangle, & \mathcal D=\NVE, \\
      |\mathbf q,\mathbf p,\xi\rangle, & \mathcal D=\NVT,
    \end{cases}
    \\
    \mathrm d\Gamma_{\mathcal D}
    &=
    \begin{cases}
      \displaystyle
      \prod_{i=1}^{N_f}\mathrm d q_i\,\mathrm d p_i,
      & \mathcal D=\NVE, \\
      \displaystyle
      \left(
        \prod_{i=1}^{N_f}\mathrm d q_i\,\mathrm d p_i
      \right)
      \mathrm d\xi,
      & \mathcal D=\NVT.
    \end{cases}
  \end{aligned}
  \label{eq:phase-space-register-definitions-D}
\end{equation}
Let $\varrho_{{\rm eq},\mathcal D}(\mathbf z_{\mathcal D})$ be the normalized
equilibrium density,
\begin{equation}
  \int
  \varrho_{{\rm eq},\mathcal D}(\mathbf z_{\mathcal D})
  \mathrm d\Gamma_{\mathcal D}
  =
  1.
  \label{eq:eq-density-normalization-D}
\end{equation}
The equilibrium KvN state is defined by
\begin{equation}
  \begin{aligned}
    \psi_{{\rm eq},\mathcal D}(\mathbf z_{\mathcal D})
    &=
    \sqrt{
      \varrho_{{\rm eq},\mathcal D}(\mathbf z_{\mathcal D})
    },
    \\
    |\psi_{{\rm eq},\mathcal D}\rangle
    &=
    \int
    \psi_{{\rm eq},\mathcal D}(\mathbf z_{\mathcal D})
    |\mathbf z_{\mathcal D}\rangle
    \mathrm d\Gamma_{\mathcal D}.
  \end{aligned}
  \label{eq:eq-kvn-state-D}
\end{equation}
On a discrete quantum register, the integral in
Eq.~\eqref{eq:eq-kvn-state-D} is replaced by the corresponding normalized
sum over grid points.

We assume a state-preparation oracle for the normalized flux-excited state
\begin{equation}
  |\alpha_{\mathcal D}\rangle
  =
  \frac{
    \hat J|\psi_{{\rm eq},\mathcal D}\rangle
  }{
    \sqrt{
      \langle\psi_{{\rm eq},\mathcal D}|
      \hat J^2
      |\psi_{{\rm eq},\mathcal D}\rangle
    }
  }.
  \label{eq:alpha-state-circuit-section}
\end{equation}
Here, $\hat J$ is the scalar flux operator used for the target transport
coefficient. For the velocity flux of the $\ell$th scalar degree of freedom,
\begin{equation}
  \hat J_\ell
  =
  \frac{\hat p_\ell}{m_\ell},
  \qquad
  \ell=1,\ldots,N_f.
  \label{eq:velocity-flux-component}
\end{equation}
A diagonal flux operator $J(\hat{\mathbf q},\hat{\mathbf p})$ can be included
by amplitude preparation based on the diagonal function or by an LCU-type
state-preparation procedure.

The controlled unitary used in QPE is
\begin{equation}
  U
  =
  S_2^{\mathcal D}(\Delta t),
  \qquad
  \mathcal D\in\{\NVE,\NVT\}.
  \label{eq:qpe-U-definition-circuit-section}
\end{equation}

\section{Dynamical Correlations}\label{sec:dynamic-correlations}

In this section, we define the equilibrium KvN wave function, the
flux-excited state, and the velocity autocorrelation function (VACF)
needed for Green--Kubo transport coefficients. The KvN inner-product
representation is not restricted to $\NVE$ dynamics. The same construction
applies to Hamiltonian $\NVE$ dynamics and to extended-phase-space
$\NVT$ dynamics with a Nos\'e--Hoover thermostat, provided that the
corresponding invariant distribution and Hermitian KvN generator are
specified. In the latter part of this section, we use a two-particle
coupled cosine system to verify that the finite-grid VACF converges to a
classical MD reference.

\subsection{Invariant Density and Equilibrium KvN Wave Function}
\label{sec:equilibrium-kvn-state}

The KvN wavefunction corresponding to
Eq.~\eqref{eq:extended-canonical-density} is chosen as the positive square
root of the extended canonical density. In the continuum notation, it
factorizes as
\begin{equation}
  \begin{aligned}
    \psi_{\rm ext}(\mathbf q,\mathbf p,\xi)
    &=
    \psi_{\rm can}(\mathbf q,\mathbf p)\,
    \chi_{\rm th}(\xi),
    \\
    \psi_{\rm can}(\mathbf q,\mathbf p)
    &=
    \frac{1}{\sqrt{Z_{\rm can}}}
    \exp
    \left[
      -\frac{\beta}{2}
      H_{\rm cl}(\mathbf q,\mathbf p)
    \right],
    \\
    \chi_{\rm th}(\xi)
    &=
    \left(
      \frac{\beta Q}{2\pi}
    \right)^{1/4}
    \exp
    \left[
      -\frac{\beta Q\xi^2}{4}
    \right].
  \end{aligned}
  \label{eq:psi-ext-factorized}
\end{equation}
Here, $\chi_{\rm th}(\xi)$ is the normalized KvN amplitude for the
Nos\'e--Hoover thermostat variable $\xi$, and
\begin{equation}
  Z_{\rm can}
  =
  \int
  \exp
  \left[
    -\beta H_{\rm cl}(\mathbf q,\mathbf p)
  \right]
  \prod_{i=1}^{N_f}
  \mathrm d q_i\,\mathrm d p_i.
  \label{eq:Zcan-definition}
\end{equation}
Thus, the extended equilibrium KvN state can be written as
\begin{equation}
  |\psi_{\rm ext}\rangle
  =
  |\psi_{\rm can}\rangle_{\mathbf q,\mathbf p}
  \otimes
  |\chi_{\rm th}\rangle_\xi.
  \label{eq:psi-ext-product}
\end{equation}

On a discrete grid, we absorb the grid weights into the normalization constant
and use
\begin{equation}
  |\psi_{\rm can}\rangle
  =
  \frac{1}{\sqrt{Z_{\rm grid}}}
  \sum_{\mathbf q,\mathbf p}
  \exp
  \left[
    -\frac{\beta}{2}
    H_{\rm cl}(\mathbf q,\mathbf p)
  \right]
  |\mathbf q,\mathbf p\rangle,
  \label{eq:canonical-amplitude-encoding}
\end{equation}
where the sum runs over all discrete grid points of the
$(\mathbf q,\mathbf p)$ registers. The normalization constant $Z_{\rm can}$ is defined by
\begin{equation}
  Z_{\rm grid}
  =
  \sum_{\mathbf q,\mathbf p}
  \exp
  \left[
    -\beta H_{\rm cl}(\mathbf q,\mathbf p)
  \right].
  \label{eq:Zgrid-definition}
\end{equation}

\subsection{KvN Inner-Product Representation of Green--Kubo Correlation Functions}
\label{sec:vacf-kvn-inner-product}

We consider the velocity autocorrelation function for the $\ell$th scalar
degree of freedom under the dynamics
$\mathcal D\in\{\NVE,\NVT\}$. The phase-space variable is
$\mathbf z_{\mathcal D}=(\mathbf q,\mathbf p)$ for $\mathcal D=\NVE$ and
$\mathbf z_{\mathcal D}=(\mathbf q,\mathbf p,\xi)$ for $\mathcal D=\NVT$.
The equilibrium expectation value of a classical observable
$A(\mathbf z_{\mathcal D})$ is
\begin{equation}
  \langle A\rangle_{{\rm eq},\mathcal D}
  =
  \int
  A(\mathbf z_{\mathcal D})
  \varrho_{{\rm eq},\mathcal D}(\mathbf z_{\mathcal D})
  \mathrm d\Gamma_{\mathcal D}.
  \label{eq:eq-expectation-D}
\end{equation}
For the corresponding multiplication operator $\hat A$, this expectation value
is equivalently written as
\begin{equation}
  \langle A\rangle_{{\rm eq},\mathcal D}
  =
  \bra{\psi_{{\rm eq},\mathcal D}}
  \hat A
  \ket{\psi_{{\rm eq},\mathcal D}},
  \label{eq:eq-expectation-kvn-D}
\end{equation}
where $|\psi_{{\rm eq},\mathcal D}\rangle$ is the equilibrium KvN state.

The velocity component and its multiplication operator are defined by
\begin{equation}
  v_\ell(\mathbf z_{\mathcal D})
  =
  \frac{p_\ell}{m_\ell},
  \qquad
  \hat v_\ell
  =
  \frac{\hat p_\ell}{m_\ell},
  \qquad
  \ell=1,\ldots,N_f.
  \label{eq:velocity-component-definition}
\end{equation}
For $\mathcal D=\NVT$, $\hat v_\ell$ acts on the physical momentum register
and as the identity on the thermostat register. The time-dependent velocity
$v_\ell(t;\mathbf z_{\mathcal D})$ denotes the value of this velocity
component at time $t$ along the classical trajectory initialized at
$\mathbf z_{\mathcal D}$. Thus,
\begin{equation}
  v_\ell(t;\mathbf z_{\mathcal D})
  =
  \frac{
    p_\ell(t;\mathbf z_{\mathcal D})
  }{
    m_\ell
  },
  \qquad
  v_\ell(0;\mathbf z_{\mathcal D})
  =
  \frac{p_\ell}{m_\ell}.
  \label{eq:velocity-time-dependent-definition}
\end{equation}

The velocity autocorrelation function is
\begin{equation}
  \begin{aligned}
    C_{vv,\ell}^{\mathcal D}(t)
    &=
    \left\langle
      v_\ell(t)v_\ell(0)
    \right\rangle_{{\rm eq},\mathcal D}
    \\
    &=
    \int
    v_\ell(t;\mathbf z_{\mathcal D})
    v_\ell(0;\mathbf z_{\mathcal D})
    \varrho_{{\rm eq},\mathcal D}(\mathbf z_{\mathcal D})
    \mathrm d\Gamma_{\mathcal D}.
  \end{aligned}
  \label{eq:vacf-definition}
\end{equation}
At $t=0$,
\begin{equation}
  C_{vv,\ell}^{\mathcal D}(0)
  =
  \left\langle
    v_\ell^2
  \right\rangle_{{\rm eq},\mathcal D}
  =
  \bra{\psi_{{\rm eq},\mathcal D}}
  \hat v_\ell^2
  \ket{\psi_{{\rm eq},\mathcal D}}.
  \label{eq:vacf-zero-time}
\end{equation}

The normalized velocity-weighted KvN state is
\begin{equation}
  |\alpha_{\mathcal D,\ell}\rangle
  =
  \frac{
    \hat v_\ell
    |\psi_{{\rm eq},\mathcal D}\rangle
  }{
    \sqrt{
      \bra{\psi_{{\rm eq},\mathcal D}}
      \hat v_\ell^2
      \ket{\psi_{{\rm eq},\mathcal D}}
    }
  }
  =
  \frac{
    \hat v_\ell
    |\psi_{{\rm eq},\mathcal D}\rangle
  }{
    \sqrt{
      C_{vv,\ell}^{\mathcal D}(0)
    }
  }.
  \label{eq:velocity-weighted-state}
\end{equation}
The KvN time-evolution operator is
\begin{equation}
  U_{\mathcal D}(t)
  =
  \exp
  \left[
    -\ii
    \hat H_{\rm KvN}^{\mathcal D}
    t
  \right].
  \label{eq:kvn-time-evolution-D}
\end{equation}
With this convention, the normalized VACF becomes the Hilbert-space
autocorrelation function
\begin{equation}
  c_{vv,\ell}^{\mathcal D}(t)
  \equiv
  \frac{
    C_{vv,\ell}^{\mathcal D}(t)
  }{
    C_{vv,\ell}^{\mathcal D}(0)
  }
  =
  \bra{\alpha_{\mathcal D,\ell}}
  U_{\mathcal D}(t)
  \ket{\alpha_{\mathcal D,\ell}}.
  \label{eq:vacf-inner-product}
\end{equation}
Depending on the sign convention used for the KvN generator, the right-hand
side of Eq.~\eqref{eq:vacf-inner-product} may instead give
$c_{vv,\ell}^{\mathcal D}(-t)$. This distinction does not affect the
Green--Kubo integral when the equilibrium VACF is real and time-reversal
symmetric.

Equation~\eqref{eq:vacf-inner-product} can be evaluated by QPE. or by a
Hadamard test with a controlled implementation of $U_{\mathcal D}(t)$. In the
Hadamard-test approach, the real part of
$\bra{\alpha_{\mathcal D,\ell}}U_{\mathcal D}(t)
\ket{\alpha_{\mathcal D,\ell}}$ gives the normalized VACF.

\subsection{Validation Setup Using a Two-Particle Coupled Cosine System}\label{sec:two-particle-cosine-test}

To validate the dynamical correlations, we use a two-particle, one-dimensional coupled cosine potential,
\begin{equation}
  V(q_1,q_2)
  =V_0\left[\cos(q_1-q_2)
  +\varepsilon(\cos q_1+\cos q_2)\right].
  \label{eq:two-particle-cosine-potential}
\end{equation}
In the numerical examples, we set \(V_0=5\), \(\varepsilon=1.2\), \(m=1\), and \(T_0=1\).

A single-particle \(\NVE\) system in a cosine potential is a one-degree-of-freedom Hamiltonian system, and its trajectories are integrable. In particular, running trajectories that cross the barrier give ballistic contributions at long times, making the single-particle model unsuitable as a validation system for transport coefficients associated with normal diffusion. With two particles, the interparticle coupling competes with the external periodic potential. This competition induces energy exchange and nonlinear mixing, promotes chaotic behavior, and produces a normal-diffusive component over finite observation times. For this reason, the present system serves as a minimal model in which the short-time oscillation of the VACF, its long-time decay, and the onset of finite-grid errors can be examined within the same setup.

Each position coordinate is represented on a periodic grid, and \(N_p=64\) grid points are used for each momentum coordinate. The momentum range is chosen sufficiently wide according to the discretization criteria discussed above. For the number \(N_q\) of position-grid points, we compare \(N_q=16,32,64,128\) for \(\NVE\) and \(N_q=16,32,48\) for \(\NVT\). For the classical MD reference, canonical initial conditions were sampled by Langevin \(\NVT\) simulations~\cite{Leimkuhler2013}, and the subsequent production runs were performed using either \(\NVE\) dynamics or single-thermostat Nos\'e--Hoover dynamics.

For the selected dynamics \({\cal D}\), the quantity compared in this section is
\begin{equation}
  \Delta c_{\cal D}(t)
  =c_{\cal D}^{\rm KvN}(t)-c_{\cal D}^{\rm MD}(t).
  \label{eq:delta-c-definition-general}
\end{equation}
Here \(c_{\cal D}^{\rm KvN}(t)\) is computed from the discrete KvN propagator through Eq.~\eqref{eq:vacf-inner-product}, whereas \(c_{\cal D}^{\rm MD}(t)\) is the VACF obtained from the corresponding average over classical trajectories. Because a finite number of MD trajectories is used, \(\Delta c_{\cal D}(t)\) contains not only the finite-grid error on the KvN side but also the statistical error of the MD reference.

For the numerical validation in this work, we applied the matrix operations
corresponding to the quantum gates by state-vector simulation using NumPy and
CuPy. As shown in Appendix~\ref{app:circuit-validation}, the resulting
propagators and QPE bin distributions agree with Qiskit~\cite{qiskit2024} statevector simulations to within machine precision.

\begin{figure*}[t]
  \centering
  \includegraphics[width=0.92\textwidth]{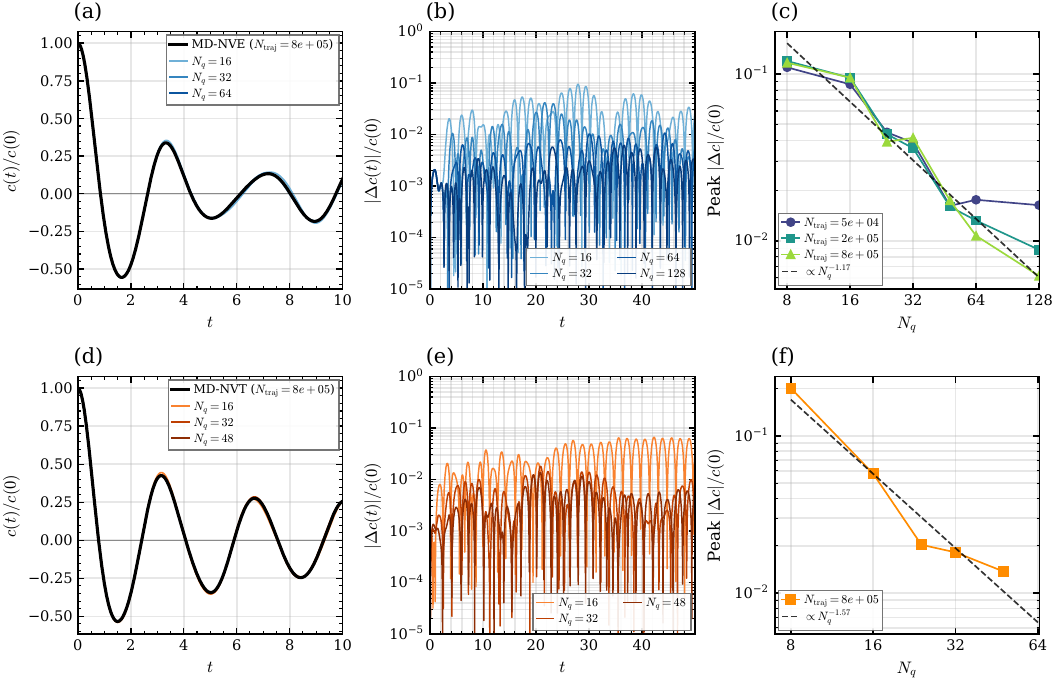}
  \caption{\textbf{VACF convergence and finite-grid effects in the two-particle coupled cosine system.}
  The upper and lower rows show the \(\NVE\) and Nos\'e--Hoover \(\NVT\) results, respectively.
  The system is defined by Eq.~\eqref{eq:two-particle-cosine-potential} with \(V_0=5\), \(\varepsilon=1.2\), \(m=T_0=1\), and \(N_p=64\).
  The left column shows the short-time VACF, the middle column shows the deviation \(|\Delta c(t)|/c(0)\) from the MD reference, and the right column shows the peak deviation as a function of \(N_q\).
  The peak deviation decreases empirically as approximately \(N_q^{-1.17}\) for \(\NVE\) and \(N_q^{-1.57}\) for \(\NVT\).
  These exponents are empirical fits for this validation system and should not be interpreted as universal exponents.}
  \label{fig:cvv-grid}
\end{figure*}

Figures~\ref{fig:cvv-grid}(a)--\ref{fig:cvv-grid}(c) show the validation for Hamiltonian \(\NVE\) evolution. Panel (a) shows the short-time VACF, panel (b) shows the time evolution of \(|\Delta c(t)|/c(0)\), and panel (c) shows the \(N_q\) dependence of the peak deviation. Figures~\ref{fig:cvv-grid}(d)--\ref{fig:cvv-grid}(f) show the same analysis for Nos\'e--Hoover \(\NVT\) evolution, where the grid convergence of \(\Cvv(t)\) is evaluated through the KvN inner-product representation on the extended phase space including the thermostat variable.

\subsection{Finite-Grid Errors and Grid Convergence}\label{sec:finite-grid-correlation-error}

Figures~\ref{fig:cvv-grid}(a) and \ref{fig:cvv-grid}(d) show the short-time VACF obtained with the \(\NVE\) and \(\NVT\) propagators, respectively.
In both cases, the curves converge as the position grid is refined, and the short-time oscillatory structure agrees with the classical MD reference.
This confirms that, under appropriate discretization conditions, the selected Liouville evolution is stably reproduced by the discrete KvN propagator.

At longer times, finite-grid effects become visible.
Figures~\ref{fig:cvv-grid}(b) and \ref{fig:cvv-grid}(e) show \(|\Delta c(t)|/c(0)\) on a logarithmic scale.
Increasing \(N_q\) delays the onset of error growth.
This behavior is analogous to filamentation, which is well known in Vlasov-type grid simulations~\cite{Cheng1976,Klimas1987,Klimas1994,Vogman2014}.
As the phase-space flow develops increasingly fine structures in the KvN wave function, Fourier components beyond the Nyquist wavenumber of the finite grid can no longer be represented accurately.
Increasing the number of grid points delays this onset, but it does not eliminate the finite-grid limitation at arbitrarily long times.

Figures~\ref{fig:cvv-grid}(c) and \ref{fig:cvv-grid}(f) show the peak deviation over \(0\leq t\leq T_{\rm sim}\) as a function of \(N_q\).
Within the present validation range, we observe the empirical scaling
\begin{equation}
  \max_{0\leq t\leq T_{\rm sim}}
  \frac{|\Delta c_{\NVE}(t)|}{c_{\NVE}(0)}
  \propto N_q^{-1.17},
  \label{eq:nve-peak-scaling}
\end{equation}
for \(\NVE\), and
\begin{equation}
  \max_{0\leq t\leq T_{\rm sim}}
  \frac{|\Delta c_{\NVT}(t)|}{c_{\NVT}(0)}
  \propto N_q^{-1.57}
  \label{eq:nvt-peak-scaling}
\end{equation}
for \(\NVT\).
When a lower-statistics MD reference is used, the error reaches a floor at large \(N_q\).
This floor is not a finite-grid error of the KvN calculation, but is caused by statistical noise in the classical trajectory average.

More generally, let \(z\) denote the grid axis that dominates the correlation-function error, and let \(N_z\) be the number of grid points along that axis.
Suppose that, over the observation range of interest, the correlation-function error follows
\begin{equation}
  \epsilon_C(N_z)
  \simeq
  A N_z^{-\gamma}.
  \label{eq:generic-grid-scaling}
\end{equation}
For a quantum register representation, \(N_z=2^{n_z}\), and therefore
\begin{equation}
  \epsilon_C(n_z)
  \simeq
  A2^{-\gamma n_z}.
\end{equation}
The number of qubits required to reach a target error \(\epsilon_C^{\rm target}\) is then estimated as
\begin{equation}
  n_z
  \gtrsim
  \frac{1}{\gamma}
  \log_2\left(\frac{A}{\epsilon_C^{\rm target}}\right).
  \label{eq:nz-qubits-for-target-error}
\end{equation}
The exponents \(\gamma=1.17\) and \(1.57\) in Fig.~\ref{fig:cvv-grid} are empirical values for the present validation system, not universal convergence exponents.
The essential point is that algebraic convergence with respect to the number of grid points \(N_z\) appears as exponential error reduction with respect to the register size \(n_z\).
At the level of grid representation, this means that increasing the register by one qubit doubles the resolution along that axis, whereas a corresponding classical grid representation must explicitly store the doubled number of grid points.

Thus, the KvN inner product of the velocity-weighted state gives the normalized VACF for the selected dynamics \({\cal D}\).
The finite-grid error in \(\Cvv(t)\) decreases approximately as \(N_z^{-\gamma}\) over the observation range, and using a quantum register with \(N_z=2^{n_z}\) translates this grid convergence into an exponential decrease with the number of qubits assigned to that axis.
In the next section, we introduce the procedure for reading out this correlation function as a Green--Kubo integral.

\section{QPE Bin-Zero Readout of Green--Kubo Integrals}\label{sec:qpe-readout}

In the preceding section, we showed that the autocorrelation of a flux-excited KvN state gives the normalized time-correlation function associated with the selected dynamics \({\cal D}\). In this section, we use this flux-excited state as the input to quantum phase estimation (QPE) and show that the bin-zero probability \(P_0\) of the ancilla register exactly yields a finite-window Bartlett-weighted Green--Kubo estimator. In Fig.~\ref{fig:D-vs-t}, we validate the Bartlett-windowed Green--Kubo estimator obtained from the QPE bin-zero readout for both \(\NVE\) dynamics and Nos\'e--Hoover-type \(\NVT\) dynamics.

\subsection{Green--Kubo Formula and the Flux-Excited State}
\label{sec:gk-kvn}

We now apply the KvN inner-product representation of
Sec.~\ref{sec:vacf-kvn-inner-product} to the Green--Kubo formula. For a scalar
flux $J(\mathbf z_{\mathcal D})$ under the selected dynamics
$\mathcal D\in\{\NVE,\NVT\}$, the equilibrium flux autocorrelation function is
\begin{equation}
  C_{JJ}^{\mathcal D}(t)
  =
  \left\langle
    J(t)J(0)
  \right\rangle_{{\rm eq},\mathcal D}.
  \label{eq:JJ-correlation-definition}
\end{equation}
Here, $J$ is understood as a mean-subtracted flux when its equilibrium mean is
nonzero. The corresponding multiplication operator is
$\hat J=J(\hat{\mathbf z}_{\mathcal D})$, and the normalized flux-excited KvN
state is
\begin{equation}
  |\alpha_{\mathcal D,J}\rangle
  =
  \frac{
    \hat J|\psi_{{\rm eq},\mathcal D}\rangle
  }{
    \sqrt{
      C_{JJ}^{\mathcal D}(0)
    }
  },
  \qquad
  C_{JJ}^{\mathcal D}(0)
  =
  \bra{\psi_{{\rm eq},\mathcal D}}
  \hat J^2
  \ket{\psi_{{\rm eq},\mathcal D}}.
  \label{eq:qpe-alpha-state}
\end{equation}
The normalized correlation function is then evaluated using the overlap
representation derived in Eq.~\eqref{eq:vacf-inner-product}.

For the diffusion coefficient considered in this work, the flux is one velocity
component. For the $\ell$th scalar degree of freedom,
\begin{equation}
  J(\mathbf z_{\mathcal D})
  =
  v_\ell(\mathbf z_{\mathcal D})
  =
  \frac{p_\ell}{m_\ell},
  \qquad
  \hat J
  =
  \hat v_\ell
  =
  \frac{\hat p_\ell}{m_\ell}.
  \label{eq:velocity-flux-as-J}
\end{equation}
Therefore,
\begin{equation}
  C_{JJ}^{\mathcal D}(t)
  =
  C_{vv,\ell}^{\mathcal D}(t),
  \qquad
  C_{JJ}^{\mathcal D}(0)
  =
  C_{vv,\ell}^{\mathcal D}(0).
  \label{eq:velocity-correlation-as-J}
\end{equation}
For a canonical distribution, the zero-time value is
\begin{equation}
  C_{vv,\ell}^{\mathcal D}(0)
  =
  \frac{T_0}{m_\ell}.
  \label{eq:canonical-velocity-variance}
\end{equation}
The one-dimensional diffusion coefficient associated with this component is
\begin{equation}
  D_{\mathcal D}
  =
  \int_0^\infty
  C_{vv,\ell}^{\mathcal D}(t)\,
  \mathrm d t.
  \label{eq:gk-diffusion}
\end{equation}
We use $\mathcal D=\NVE$ for Hamiltonian production dynamics and
$\mathcal D=\NVT$ when the correlation function of the thermostatted dynamics
itself is considered.

\subsection{QPE Bin-Zero Probability and the Bartlett Window}
\label{sec:qpe-bartlett}

The one-step propagator used in QPE is
\begin{equation}
  U_{\mathcal D}
  =
  U_{\mathcal D}(\Delta t),
  \label{eq:qpe-one-step-propagator}
\end{equation}
where $\Delta t$ is the discrete time step. The system register is initialized
in the flux-excited state $|\alpha_{\mathcal D,J}\rangle$ defined in
Eq.~\eqref{eq:qpe-alpha-state}. With $m_{\rm anc}$ ancilla qubits, the number
of QPE time samples is
\begin{equation}
  K
  =
  2^{m_{\rm anc}},
  \qquad
  r
  =
  0,1,\ldots,K-1.
  \label{eq:qpe-K-definition}
\end{equation}
QPE combines the powers $U_{\mathcal D}^{r}$ through controlled applications
of $U_{\mathcal D}^{2^j}$. The corresponding finite time window is
\begin{equation}
  \tau_{\QPE}
  =
  K\Delta t
  =
  2^{m_{\rm anc}}\Delta t.
  \label{eq:qpe-time-window}
\end{equation}
In this subsection, $P_0$ denotes the probability that the $m_{\rm anc}$-qubit
QPE ancilla register is measured in the zero bin, namely in the all-zero
computational-basis state $\ket{0}^{\otimes m_{\rm anc}}$, when the system
register is initialized in $|\alpha_{\mathcal D,J}\rangle$ and the controlled
unitary is $U_{\mathcal D}$.

The largest sampled time is $(K-1)\Delta t$, while the Bartlett window below
vanishes at $K\Delta t$. The frequency resolution near zero phase is therefore
estimated as
\begin{equation}
  \delta\omega
  \sim
  \frac{2\pi}{\tau_{\QPE}}.
  \label{eq:qpe-frequency-resolution}
\end{equation}
Increasing $m_{\rm anc}$ improves this resolution, but also increases the
largest controlled power of $U_{\mathcal D}$.

We denote the eigenstates and eigenphases of $U_{\mathcal D}$ by
$|\phi_k\rangle$ and $\theta_k$, respectively:
\begin{equation}
  U_{\mathcal D}|\phi_k\rangle
  =
  e^{-\ii\theta_k}
  |\phi_k\rangle.
  \label{eq:qpe-eigenphase-definition}
\end{equation}
The input state is expanded in this eigenbasis as
\begin{equation}
  |\alpha_{\mathcal D,J}\rangle
  =
  \sum_k
  a_k|\phi_k\rangle,
  \qquad
  a_k
  =
  \langle\phi_k|\alpha_{\mathcal D,J}\rangle.
  \label{eq:qpe-input-eigenbasis-expansion}
\end{equation}
Here, $a_k$ is the expansion coefficient of the flux-excited state in the
eigenbasis of $U_{\mathcal D}$. Standard QPE gives the zero-bin probability as
\begin{equation}
  P_0
  =
  \sum_k
  |a_k|^2
  \left|
    \frac{1}{K}
    \sum_{r=0}^{K-1}
    e^{-\ii r\theta_k}
  \right|^2.
  \label{eq:qpe-p0-fejer}
\end{equation}
This is the eigenphase distribution of $|\alpha_{\mathcal D,J}\rangle$
averaged around zero phase by the Fej\'er kernel.

Using the squared-Dirichlet-kernel identity
\begin{equation}
  \left|
    \sum_{r=0}^{K-1}
    e^{-\ii r\theta}
  \right|^2
  =
  K
  +
  2
  \sum_{s=1}^{K-1}
  (K-s)\cos(s\theta),
  \label{eq:dirichlet-square-identity}
\end{equation}
and the normalized correlation function
\begin{equation}
  c_{JJ}^{\mathcal D}(s\Delta t)
  =
  \mathrm{Re}
  \bra{\alpha_{\mathcal D,J}}
  U_{\mathcal D}^{s}
  \ket{\alpha_{\mathcal D,J}},
  \label{eq:normalized-correlation-discrete}
\end{equation}
we obtain
\begin{equation}
  P_0
  =
  \frac{1}{K^2}
  \left[
    K
    +
    2
    \sum_{s=1}^{K-1}
    (K-s)
    c_{JJ}^{\mathcal D}(s\Delta t)
  \right].
  \label{eq:P0-bartlett}
\end{equation}
Equation~\eqref{eq:dirichlet-square-identity} follows from the
self-convolution of a finite geometric series; the derivation is given in
Appendix~\ref{app:qpe-bartlett}.

Multiplying Eq.~\eqref{eq:P0-bartlett} by
$C_{JJ}^{\mathcal D}(0)\tau_{\QPE}/2$ gives the Bartlett-windowed
Green--Kubo integral
\begin{equation}
  \begin{aligned}
    \mathcal I_{\rm Bart}^{\mathcal D,J}(K)
    &\equiv
    \frac{1}{2}
    C_{JJ}^{\mathcal D}(0)
    \tau_{\QPE}
    P_0
    \\
    &=
    C_{JJ}^{\mathcal D}(0)
    \Delta t
    \left[
      \frac{1}{2}
      +
      \sum_{s=1}^{K-1}
      \left(
        1-\frac{s}{K}
      \right)
      c_{JJ}^{\mathcal D}(s\Delta t)
    \right].
  \end{aligned}
  \label{eq:D-bartlett-qpe}
\end{equation}
The factor
\begin{equation}
  w_s
  =
  1-\frac{s}{K},
  \qquad
  s=0,1,\ldots,K,
  \label{eq:bartlett-window-discrete}
\end{equation}
is the triangular Bartlett window. It vanishes at $s=K$, and the coefficient
of the $s=0$ term is $1/2$. Omitting this endpoint coefficient breaks the
correspondence with the QPE zero-bin probability and leaves an
$\bigO(\Delta t)$ systematic offset.

For diffusion, $J=v_\ell$ and
$C_{JJ}^{\mathcal D}(t)=C_{vv,\ell}^{\mathcal D}(t)$. The estimator used in
this work is therefore
\begin{equation}
  D_{\rm Bart}^{\mathcal D,\ell}(K)
  =
  \frac{1}{2}
  C_{vv,\ell}^{\mathcal D}(0)
  \tau_{\QPE}
  P_0,
  \qquad
  \tau_{\QPE}
  =
  2^{m_{\rm anc}}\Delta t.
  \label{eq:qpe-bin0-main-formula}
\end{equation}
Thus, the QPE readout does not require reconstructing the full phase
distribution. It is sufficient to estimate the zero-bin probability $P_0$ for
the QPE circuit initialized with the velocity-excited state.

\subsection{Finite-Time Window and Finite-Grid Error}
\label{sec:qpe-errors}

Equation~\eqref{eq:qpe-bin0-main-formula} gives the exact finite-window
Bartlett estimator associated with the QPE time window $\tau_{\QPE}$. To recover
the physical diffusion coefficient $D_{\mathcal D}$, the window length must be
long compared with the decay time of the velocity autocorrelation function. In
continuous time, the corresponding Bartlett-windowed integral is
\begin{equation}
  D_{\rm Bart}^{\mathcal D,\ell}(\tau)
  \simeq
  \int_0^{\tau}
  \left(
    1-\frac{t}{\tau}
  \right)
  C_{vv,\ell}^{\mathcal D}(t)\,
  \mathrm d t.
  \label{eq:continuous-bartlett}
\end{equation}
The window factor suppresses spectral leakage, but at finite $\tau$ it also
attenuates long-time components of the correlation function. Therefore,
$D_{\rm Bart}^{\mathcal D,\ell}(\tau)$ retains a finite-window bias. When
$C_{vv,\ell}^{\mathcal D}(t)$ decays sufficiently rapidly and $\tau$ exceeds
the dominant correlation time, the leading correction is expected to scale as
$1/\tau$, or equivalently as $1/K$. We model the convergence regime as
\begin{equation}
  D_{\rm Bart}^{\mathcal D,\ell}(K)
  =
  D_\infty^{\mathcal D,\ell}
  +
  \frac{a_1}{K}
  +
  \mathcal O(K^{-2}),
  \label{eq:richardson-form}
\end{equation}
and estimate $D_\infty^{\mathcal D,\ell}$ by extrapolation in $1/K$.

In Sec.~\ref{sec:finite-grid-correlation-error}, the finite-grid error of the
correlation function within a fixed observation window was found to decrease
approximately as $N_z^{-\gamma}$. This error propagates linearly to the
Bartlett estimator. We denote by
$c_{vv,\ell}^{\mathcal D,{\rm disc}}(t)$ the normalized VACF obtained from the
finite-grid discrete KvN calculation, and by
$c_{vv,\ell}^{\mathcal D,{\rm cont}}(t)$ the corresponding normalized VACF of
the continuum dynamics. At the discrete time $t=s\Delta t$, define
\begin{equation}
  \Delta c_s
  =
  c_{vv,\ell}^{\mathcal D,{\rm disc}}(s\Delta t)
  -
  c_{vv,\ell}^{\mathcal D,{\rm cont}}(s\Delta t).
  \label{eq:discrete-continuum-correlation-error}
\end{equation}
If $|\Delta c_s|\leq \epsilon_c$ for $s=0,1,\ldots,K-1$, then
Eq.~\eqref{eq:D-bartlett-qpe} gives
\begin{equation}
  \begin{aligned}
    |\Delta D_{\rm Bart}^{\mathcal D,\ell}|
    &\leq
    C_{vv,\ell}^{\mathcal D}(0)
    \Delta t
    \left[
      \frac{1}{2}
      +
      \sum_{s=1}^{K-1}
      \left(
        1-\frac{s}{K}
      \right)
    \right]
    \epsilon_c
    \\
    &=
    \frac{1}{2}
    C_{vv,\ell}^{\mathcal D}(0)
    \tau_{\QPE}
    \epsilon_c .
  \end{aligned}
  \label{eq:cvv-error-to-d-bart}
\end{equation}
This estimate gives the accuracy required for the normalized correlation
function to achieve a target error in $D_{\rm Bart}^{\mathcal D,\ell}$.
Together with Eq.~\eqref{eq:nz-qubits-for-target-error}, it also estimates the
number of grid qubits when the grid size is written as $N_z=2^{n_z}$.

The systematic error in $D_{\rm Bart}^{\mathcal D,\ell}$ mainly consists of
finite-grid error, Trotter error, and finite-window bias. The statistical error
from estimating $P_0$ with a finite number of shots is treated separately in
the circuit-validation and amplitude-estimation analyses. In particular, MLAE
is introduced to improve the query scaling for estimating $P_0$ relative to
direct shot sampling.

\subsection{Numerical Validation for the Two-Particle Coupled Cosine System}\label{sec:D-vs-t}

\begin{figure}[t]
  \centering
  \includegraphics[width=\linewidth]{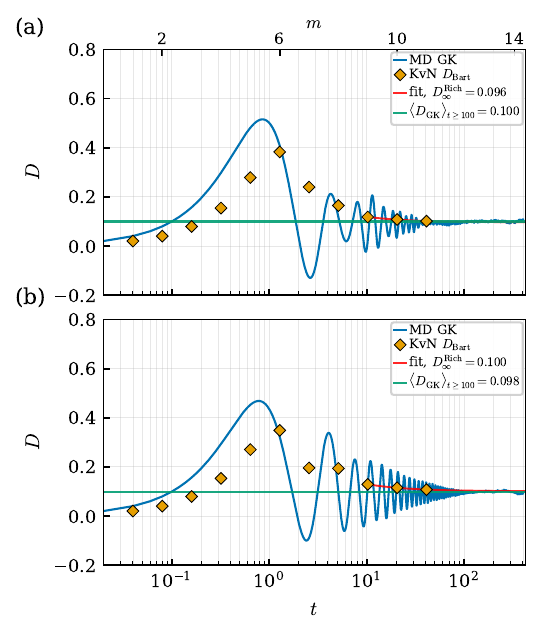}
  \caption{\textbf{Comparison between the Green--Kubo integral and the KvN Bartlett estimator.}
  (a) \(\NVE\) dynamics with \(N_q=128\).
  (b) Nos\'e--Hoover-type \(\NVT\) dynamics with \(N_q=48\).
  The blue curve shows the cumulative Green--Kubo integral
\(D_{\rm GK}(t)=\int_0^t C_{vv}(t')\,\mathrm{d}t'\)
obtained from the classical MD velocity autocorrelation function,
while the yellow diamonds show the KvN Bartlett estimator \(\Dbart(m_{\rm anc})\).
  The lower axis gives the physical time \(t\), and the upper axis gives
  \(m=\log_2(t/\Delta t)\).
  The red line shows a first-order extrapolation in \(1/K\) using the
  \(m=9,10,11\) points, and the green line shows the classical long-time
  reference \(\langle D_{\rm GK}\rangle_{100\le t\le400}\).}
  \label{fig:D-vs-t}
\end{figure}

Figure~\ref{fig:D-vs-t} shows a numerical validation of the QPE bin-zero
readout for the two-particle coupled cosine system. We apply the same
readout formula to both \(\NVE\) dynamics and Nos\'e--Hoover-type
\(\NVT\) dynamics. The blue curve is the finite-time Green--Kubo integral
\(D_{\rm GK}(t)\), obtained by directly integrating the velocity
autocorrelation function from classical MD. The yellow diamonds are
\(\Dbart(m_{\rm anc})\), computed from KvN state-vector propagation. By
Eq.~\eqref{eq:qpe-bin0-main-formula}, each marker is the bin-zero readout
value that would be returned, in the infinite-shot limit, by a QPE circuit
using the corresponding \(U_{\cal D}(\Delta t)\) and flux-excited state
\(\ket{\alpha_{\cal D}}\).

The finite-\(m_{\rm anc}\) value \(\Dbart(m_{\rm anc})\) is not simply the finite-time
Green--Kubo integral sampled at \(t=\tau_{\QPE}=2^{m_{\rm anc}}\Delta t\). As shown in
Eq.~\eqref{eq:D-bartlett-qpe}, it is a discrete integral of the
correlation function over \(0\leq t\leq \tau_{\QPE}\), multiplied by the
Bartlett window. Because the correlation function in the present system
contains oscillatory components, a large transient peak appears for short
window lengths, and the estimator returns toward its long-time value as
\(m_{\rm anc}\) increases. This nonmonotonic behavior is not a failure of QPE, but a
feature of the finite-window Bartlett estimator.

The classical finite-time Green--Kubo integral does not become perfectly
constant for finite trajectory lengths. We therefore use the average of
\(D_{\rm GK}(t)\) over the interval \(100\le t\le400\) as the classical
long-time reference. For \(\NVE\), we obtain
\begin{equation}
  \langle D_{\rm GK}^{\NVE}\rangle_{100\le t\le400}
  =0.1000\pm0.0035,
  \label{eq:D-gk-late-nve}
\end{equation}
and a first-order extrapolation in \(1/K\) using the \(\Dbart(m)\) values
at \(m=9,10,11\) gives
\begin{equation}
  D_{\infty,\NVE}^{\rm Rich}
  =0.0960\pm0.0012.
  \label{eq:D-rich-nve}
\end{equation}
For \(\NVT\), the same analysis gives
\begin{align}
  \langle D_{\rm GK}^{\NVT}\rangle_{100\le t\le400}
  &=0.0976\pm0.0043,\label{eq:D-gk-late-nvt}\\
  D_{\infty,\NVT}^{\rm Rich}
  &=0.1004\pm0.0006.\label{eq:D-rich-nvt}
\end{align}
The uncertainties quoted here are either the fluctuations of the classical
MD running integral within the long-time window or the standard errors of
the Richardson extrapolation. For both dynamics, the extrapolated KvN
Bartlett estimator agrees with the long-time Green--Kubo reference from
classical MD within a few percent.

This comparison verifies two points. First, the QPE bin-zero probability
obtained from the flux-excited state returns the Bartlett-windowed
Green--Kubo estimator defined by Eq.~\eqref{eq:qpe-bin0-main-formula}.
Second, in the regime where the finite-window bias is well described by
the leading \(1/K\) correction, extrapolating \(\Dbart(m_{\rm anc})\) in \(1/K\)
gives a value consistent with the long-time Green--Kubo integral from
classical MD. The remaining accuracy is controlled by finite-window
extrapolation, finite-grid error, Trotter error, and the statistical error
of the classical reference.

In summary, the quantum readout of the Green--Kubo integral proceeds as
follows. The correlation function is represented as the autocorrelation of
the flux-excited state \(\ket{\alpha_{\cal D}}\) under the KvN propagator
\(U_{\cal D}\). Supplying this state to QPE gives a bin-zero probability
\(P_0\), which corresponds to the discrete Green--Kubo integral multiplied
by a Bartlett window. The resulting finite-\(K\) bias can be corrected by
a \(1/K\) extrapolation in the convergence regime.

\section{\texorpdfstring{Quadratic Speedup in $P_0$ Estimation via Amplitude Estimation}{Quadratic Speedup in P0 Estimation via Amplitude Estimation}}\label{sec:amplitude-estimation}

In Sec.~\ref{sec:qpe-readout}, we showed that the flux-excited state
\(\ket{\alpha_{\cal D}}\), together with the KvN propagator
\(U=S_2^{\cal D}(\Delta t)\), maps the Bartlett-windowed Green--Kubo
estimator to the QPE bin-zero probability \(P_0\). In this section, we use amplitude estimation instead of direct shot sampling to estimate \(P_0\). For the
fixed QPE oracle defining \(P_0\), this improves the query complexity of
statistical estimation from \(O(\epsilon^{-2})\) to \(O(\epsilon^{-1})\).

\begin{figure}[t]
  \centering
  \includegraphics[width=\linewidth]{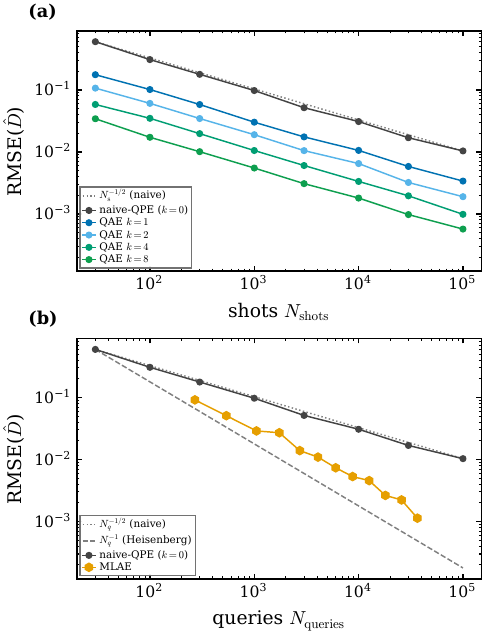}
  \caption{\textbf{RMSE reduction by amplitude estimation for the QPE bin-zero readout.}
  The calculation uses the two-particle coupled cosine system with an
  8-qubit system register and \(m_{\rm anc}=12\) QPE ancilla qubits.
  The parameters are \(K=2^{12}\), \(\Delta t=0.02\),
  \(\tau_{\QPE}=81.92\), \(\Cvv(0)=1.1334\), and
  \(P_0^{\rm exact}=4.097\times 10^{-3}\); the state-vector reference is
  \(\Dbart=0.19020\).
  (a) Prefactor reduction for fixed Grover powers
  \(k\in\{0,1,2,4,8\}\).
  The horizontal axis is the number of shots \(N_{\rm shots}\) used for
  each fixed-\(k\) circuit, and the vertical axis is the RMSE of
  \(\hat D\) over \(N_{\rm seeds}=1000\) Monte Carlo seeds.
  All series follow \(N_{\rm shots}^{-1/2}\) scaling, while the prefactor
  decreases as \(k\) increases.
  (b) Query-count dependence of growing-schedule MLAE based on the EIS.
  The horizontal axis is
  \(N_{\rm queries}=\sum_\ell N_\ell(2k_\ell+1)\), and the vertical axis
  is the same RMSE.
  Naive QPE follows \(N_{\rm queries}^{-1/2}\) scaling, whereas the
  late-stage MLAE data have a fitted slope of \(-0.953\), close to
  \(N_{\rm queries}^{-1}\).}
  \label{fig:ae-advantage}
\end{figure}

\subsection{Statistical Error of Direct Sampling}\label{sec:direct-sampling-limit}

Let \(P_0\) be the probability of obtaining bin zero in the ancilla
measurement of an \(m_{\rm anc}\)-qubit QPE circuit. If the QPE circuit is
executed \(N_{\rm shots}\) times and bin zero is observed \(N_0\) times,
the direct-sampling estimator is
\begin{equation}
  \hat P_0=\frac{N_0}{N_{\rm shots}} .
  \label{eq:p0-direct-estimator}
\end{equation}
Each shot is a Bernoulli trial, and hence
\begin{equation}
  {\rm Var}(\hat P_0)
  =\frac{P_0(1-P_0)}{N_{\rm shots}} .
  \label{eq:p0-direct-variance}
\end{equation}
This fluctuation propagates to the statistical error of \(\hat D\) through
\begin{equation}
  \hat D
  =\frac{1}{2}C_{JJ}^{({\cal D})}(0)\tau_{\QPE}\hat P_0 .
  \label{eq:D-from-p0-estimator}
\end{equation}
For \(J=v\), Eq.~\eqref{eq:D-from-p0-estimator} is the
velocity-autocorrelation readout formula for the diffusion coefficient.

To estimate \(P_0\) with relative error \(\epsilon_{\rm rel}\), 
Eq.~\eqref{eq:p0-direct-variance} gives, approximately,
\begin{equation}
  N_{\rm shots}
  \simeq
  \frac{1-P_0}{\epsilon_{\rm rel}^2 P_0}.
  \label{eq:direct-shot-complexity}
\end{equation}
For \(P_0\ll1\), the required number of shots grows in proportion to
\(1/P_0\). This is the statistical bottleneck that arises in the readout
stage of the QPE bin-zero method.

\subsection{Oracle Structure for Amplitude Estimation}\label{sec:ae-oracle}

We regard the full QPE circuit as a single unitary \(A\). This unitary
includes preparation of the flux-excited state \(\ket{\alpha_{\cal D}}\),
the controlled-\(U^{2^j}\) operations, and the inverse QFT on the ancilla
register. Defining the ancilla bin-zero subspace as the good subspace, we
can write
\begin{equation}
  A\ket{0}
  =
  \sqrt{P_0}\ket{\Psi_{\rm good}}
  +
  \sqrt{1-P_0}\ket{\Psi_{\rm bad}} .
  \label{eq:ae-state-decomposition}
\end{equation}
Here \(\ket{0}\) denotes the all-zero input to the state-preparation and
QPE circuit, \(\ket{\Psi_{\rm good}}\) is the normalized state projected
onto the ancilla bin-zero subspace, and \(\ket{\Psi_{\rm bad}}\) belongs
to the orthogonal complement. We define the amplitude angle \(\theta\) by
\begin{equation}
  P_0=\sin^2\theta .
  \label{eq:p0-theta-definition}
\end{equation}

Standard amplitude amplification constructs the Grover iterate
\begin{equation}
  Q=-A S_0 A^\dagger S_\chi
  \label{eq:grover-iterate}
\end{equation}
from the reflection \(S_\chi\) about the good subspace and the reflection
\(S_0\) about the all-zero input state~\cite{BrassardHoyer2002}. After
applying \(Q^kA\) to \(\ket{0}\), the probability of observing a good
outcome is
\begin{equation}
  p_k(\theta)
  =
  \sin^2\left[(2k+1)\theta\right].
  \label{eq:grover-amplified-probability}
\end{equation}
For small \(P_0\), one has \(\theta\simeq\sqrt{P_0}\). The Grover power
\(k\) therefore magnifies the amplitude angle by a factor \(2k+1\) before
measurement.

\subsection{Prefactor Reduction with a Fixed Grover Power}\label{sec:fixed-k-qae}

We first consider the case in which a single Grover power \(k\) is fixed.
At a fixed \(k\), the measurement outcome is still a Bernoulli random
variable, and the RMSE therefore retains the same \(N_{\rm shots}^{-1/2}\)
scaling as direct sampling. The effect of amplitude amplification appears
instead in the prefactor.

The Fisher information per shot of the Bernoulli distribution
\(p_k(\theta)\), with respect to the amplitude angle \(\theta\), is
\begin{equation}
  I_k(\theta)
  =
  \frac{1}{p_k(1-p_k)}
  \left( \frac{\partial p_k}{\partial \theta} \right)^2
  =4(2k+1)^2.
  \label{eq:fixed-k-fisher-information}
\end{equation}
Thus, on a branch where the Grover oscillation is not folded and
\(\theta\) can be recovered uniquely, the measured probability
\(\hat p_k\) can be inverted as
\begin{equation}
  \hat\theta
  =
  \frac{1}{2k+1}\arcsin\sqrt{\hat p_k}.
  \label{eq:fixed-k-theta-estimator}
\end{equation}
This gives
\begin{equation}
  \sigma(\hat\theta)
  \simeq
  \frac{1}{2(2k+1)\sqrt{N_{\rm shots}}}.
  \label{eq:fixed-k-theta-error}
\end{equation}
Propagating this error to
\(\hat D=(1/2)\Cvv(0)\tau_{\QPE}\sin^2\hat\theta\) yields
\begin{equation}
  \sigma(\hat D)
  \simeq
  \frac{\Cvv(0)\tau_{\QPE}\sin\theta\cos\theta}
       {2(2k+1)\sqrt{N_{\rm shots}}}.
  \label{eq:fixed-k-D-error}
\end{equation}
Therefore, amplitude amplification with a fixed \(k\) can reduce the RMSE
prefactor by approximately a factor of \(2k+1\), while leaving the
\(N_{\rm shots}^{-1/2}\) scaling unchanged.

The RMSE in Fig.~\ref{fig:ae-advantage}(a) is defined using the
state-vector value of \(\Dbart\) as the reference. Specifically, for each
\(k\in\{0,1,2,4,8\}\), each shot number \(N_{\rm shots}\), and each
Monte Carlo seed \(s=1,\ldots,N_{\rm seeds}\), we estimate
\(\hat\theta^{(s)}\) from the measurement outcome using
Eq.~\eqref{eq:fixed-k-theta-estimator}. We then compute
\begin{equation}
  \hat P_0^{(s)}=\sin^2\hat\theta^{(s)},
  \qquad
  \hat D^{(s)}=\frac{1}{2}\Cvv(0)\tau_{\QPE}\hat P_0^{(s)},
  \label{eq:fixed-k-Dhat}
\end{equation}
and evaluate
\begin{equation}
  {\rm RMSE}(\hat D)
  =
  \sqrt{
  \frac{1}{N_{\rm seeds}}
  \sum_{s=1}^{N_{\rm seeds}}
  \left(\hat D^{(s)}-\Dbart\right)^2
  }.
  \label{eq:rmse-definition-ae}
\end{equation}
In the present calculation, \(N_{\rm seeds}=1000\). The reference value
\(\Dbart\) is the Bartlett estimator for the QPE window with
\(m_{\rm anc}=12\).

The fixed-\(k\) method reduces the prefactor for estimating \(P_0\) at a
single circuit depth. However, if \(k\) is made too large, the periodicity
of \(\sin^2[(2k+1)\theta]\) makes the estimate of \(\theta\) nonunique.
To avoid this branch ambiguity, one must combine measurements at multiple
Grover powers.

\subsection{Heisenberg-Type Scaling with MLAE}\label{sec:mlae-scaling}

In maximum-likelihood amplitude estimation (MLAE), circuits
\(Q^{k_\ell}A\) are executed for multiple Grover powers \(\{k_\ell\}\),
and all measurement outcomes are combined into a single likelihood
function~\cite{Suzuki2020}. If \(h_\ell\) good outcomes are obtained out
of \(N_\ell\) shots, the likelihood is
\begin{equation}
  {\cal L}(\theta)
  =
  \prod_\ell
  \left[p_{k_\ell}(\theta)\right]^{h_\ell}
  \left[1-p_{k_\ell}(\theta)\right]^{N_\ell-h_\ell},
  \label{eq:mlae-likelihood}
\end{equation}
where \(p_{k_\ell}(\theta)=\sin^2[(2k_\ell+1)\theta]\). From the
maximum-likelihood estimate \(\hat\theta\), we obtain
\(\hat P_0=\sin^2\hat\theta\) and reconstruct \(\hat D\) using
Eq.~\eqref{eq:D-from-p0-estimator}.

Following the Exponential Incremental Sequence (EIS) of Suzuki
\textit{et al.}~\cite{Suzuki2020}, we use a hierarchical schedule in
which the set of Grover powers is enlarged step by step. In the EIS, the
measurement magnification factor
\(M_\ell=2k_\ell+1\), denoted by \(m_\ell\) in the notation of Suzuki
\textit{et al.}, increases approximately as \(M_\ell\sim 2^{\ell-1}\).
We implement this sequence as nested sets
\(\mathcal K^{(L)}=\{k_\ell\}_{\ell=1}^{L}\) satisfying
\(\mathcal K^{(L+1)}\supset\mathcal K^{(L)}\), and refer to this nested
structure as a \emph{growing schedule}. Figure~\ref{fig:ae-advantage}(b)
uses this growing schedule. For each \(k_\ell\), we use only
\(N_\ell=30\) shots and evaluate the RMSE in
Eq.~\eqref{eq:rmse-definition-ae} over \(N_{\rm seeds}=1000\) Monte
Carlo seeds. Since the circuit with Grover power \(k_\ell\) contains
\(A\) a total of \(2k_\ell+1\) times, we define the total query count as
\begin{equation}
  N_{\rm queries}
  =
  \sum_\ell N_\ell(2k_\ell+1).
  \label{eq:mlae-query-count}
\end{equation}
For direct sampling, the RMSE decreases as
\(N_{\rm queries}^{-1/2}\). In contrast, MLAE with the growing schedule
can approach \(N_{\rm queries}^{-1}\) scaling. This is the standard
quadratic improvement of amplitude estimation, and iterative amplitude
estimation achieves the same query-complexity improvement~\cite{Grinko2021}.

In Fig.~\ref{fig:ae-advantage}(b), naive QPE follows the
\(N_{\rm queries}^{-1/2}\) slope, whereas the late-stage MLAE data give a
log--log regression slope of \(-0.953\). At the final stage,
\(N_{\rm queries}=36600\) and
\({\rm RMSE}\simeq 1.02\times 10^{-3}\), and
\({\rm RMSE}\times N_{\rm queries}\) remains nearly constant. This
behavior is consistent with near-Heisenberg
\(N_{\rm queries}^{-1}\) scaling in the numerical regime studied here.
Thus, for the task of estimating \(P_0\) defined by the same QPE oracle,
amplitude estimation provides a quadratic query-complexity advantage over
direct shot sampling.

\section{Resource Analysis}\label{sec:resource-limitations}

In this section, we summarize the gate costs of the KvN circuits constructed in Sec.~\ref{sec:circuit-implementation}. Our aim is to compare the $\NVE$ and $\NVT$ implementations on the same footing and to identify which circuit components can be implemented with polynomial gate cost and which become bottlenecks in the present construction. As a cost metric, we use the number of CX gates after transpiling the Qiskit circuits into the $\{\mathrm{CX},u_3\}$ basis. The CX counts reported here are logical-level measures of circuit complexity and do not include hardware noise, routing overhead specific to a device topology, or error-correction overhead.

In the following, we denote the numbers of qubits in the position, momentum, and Nos\'e--Hoover variable registers by $n_q$, $n_p$, and $n_\xi$, respectively. For $\NVE$, the total register size is $n=n_q+n_p$, whereas for $\NVT$ it is $n=n_q+n_p+n_\xi$.

\subsection{Implementation Cost of the \texorpdfstring{$\Hthree$}{H3} Dilation Block}\label{sec:h3-pauli}

The main resource difference between $\NVT$ and $\NVE$ arises from the Nos\'e--Hoover friction term,
\begin{equation}
  \Hthree=-\hat\xi\otimes \hat D_p,
  \qquad
  \hat D_p=\frac{1}{2}(\hat p\hat\lambda_p+\hat\lambda_p\hat p).
  \label{eq:h3-resource-definition}
\end{equation}
This term is specific to the thermostat. Although $\hat D_p$ is the generator of momentum-space dilations in the continuum limit, a quantum circuit must implement a finite-dimensional representation of it on a discretized momentum grid. In this work, we discretize the derivative along the $p$ axis by a centered difference and decompose the resulting finite-dimensional matrix into a real linear combination of Pauli strings. We refer to this implementation as the centered-difference Pauli-decomposition method.

\begin{figure}[t]
  \centering
  \includegraphics[width=\linewidth]{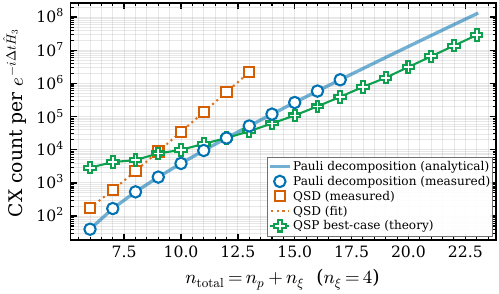}
  \caption{\textbf{CX cost of the $\Hthree$ dilation block.}
  The horizontal axis is $n_{\rm total}=n_p+n_\xi$, with $n_\xi=4$ fixed.
  The blue solid line is the analytic prediction for the centered-difference Pauli-decomposition method and follows the same $n_p2^{n_p}$-type growth as the weighted Pauli sum in Eq.~\eqref{eq:W-Dp}.
  The blue circles are the CX counts obtained by transpiling the corresponding Pauli-evolution circuits in Qiskit.
  The orange squares are the CX counts obtained by treating $e^{-\ii\delta t\Hthree}$ as a generic dense unitary and synthesizing it by quantum Shannon decomposition (QSD).
  The orange dotted line is a fit of the form $c_1 4^{n_{\rm total}}$, representing the cost of generic unitary synthesis without exploiting the structure of $\Hthree$.
  The green points show a best-case theoretical reference based on QSP Hamiltonian simulation, assuming that the sparse-access oracles described in the text are available at low cost.}
  \label{fig:h3-cost}
\end{figure}

The operator $\hat D_p$ obtained from the centered difference has a nearest-neighbor finite-difference structure. However, after binary encoding of the $p$ register, the number of Pauli strings grows with the number of momentum grid points, $N_p=2^{n_p}$. The centered-difference Pauli-decomposition results in Fig.~\ref{fig:h3-cost} are based on representing the derivative operator $\hat D_p$ by a cyclic central-difference matrix with periodic boundary conditions. Expanding this matrix in the Pauli basis on $n_p$ qubits gives a set of nonzero Pauli terms whose cardinality we denote by $N_{D_p}^{\rm cyc}(n_p)$. We also define $W_{D_p}^{\rm cyc}(n_p)$ as the corresponding weighted Pauli sum, obtained by summing the implementation weights associated with the individual Pauli-string lengths. This gives
\begin{align}
  N_{D_p}^{\rm cyc}(n_p)
  &=\frac{7}{4}2^{n_p}-(n_p+2),
  \label{eq:N-Dp}\\
  W_{D_p}^{\rm cyc}(n_p)
  &=\frac{7}{4}n_p\,2^{n_p}-3\cdot 2^{n_p}+3.
  \label{eq:W-Dp}
\end{align}
The coordinate operator $\hat\xi$ is diagonal on the $\xi$ register and can be expanded in Pauli-$Z$ strings on $n_\xi$ qubits. Therefore, the Pauli-evolution cost of $\Hthree$ is approximately proportional to $n_\xi W_{D_p}^{\rm cyc}(n_p)$. For the centered-difference Pauli-decomposition method, this gives
\begin{equation}
  N_{\rm CX}(e^{-\ii\delta t\Hthree})
  =\bigO(n_\xi n_p\,2^{n_p}).
  \label{eq:h3-pauli-asymptotic}
\end{equation}
Equivalently, since $N_p=2^{n_p}$, this scaling is quasi-linear in the number of momentum grid points, $\bigO(n_\xi N_p\log N_p)$, and therefore exponential in the number of momentum qubits $n_p$.

Figure~\ref{fig:h3-cost} compares three estimates for implementing $e^{-\ii\delta t\Hthree}$. The first is the centered-difference Pauli-decomposition method. The blue solid line gives the analytic prediction for centered-difference Pauli evolution, and the blue circles show the CX counts after Qiskit transpilation. Over the measured range, both exhibit the same $n_p2^{n_p}$-type growth. Equation~\eqref{eq:W-Dp} gives the closed form of the weighted Pauli sum responsible for this scaling. This is the implementation used in the numerical circuits of the present work.

The second estimate is dense-unitary synthesis by QSD. Here, the matrix $e^{-\ii\delta t\Hthree}$ is treated as a generic $2^{n_{\rm total}}\times2^{n_{\rm total}}$ unitary, without using the sparsity or tensor-product structure of $\Hthree$. Since the synthesis cost of a generic unitary scales as $4^{n_{\rm total}}$~\cite{ShendeBullockMarkov2006}, the orange dotted line in Fig.~\ref{fig:h3-cost} has the same exponential form. The centered-difference Pauli-decomposition method lies below QSD because it explicitly exploits the nearest-neighbor finite-difference structure of $\hat D_p$ through a Pauli-sum representation.

The third estimate is a best-case theoretical reference based on QSP~\cite{Childs2012,LowChuang2019}. For this curve, we assume that a location oracle $O_{\rm loc}$ and a value oracle $O_{\rm val}$ for block encoding $\Hthree$ as a sparse Hamiltonian are available at low cost. Thus, the green points are not transpiled counts from an implemented circuit, but target values in the sparse-access oracle model. Under this assumption, the QSP polynomial degree scales as $K_{\rm QSP}=O(2^{n_p})$, and the leading CX scaling becomes $\bigO(n_{\rm total}2^{n_p})$. The evaluation method and oracle assumptions are summarized in Appendix~\ref{app:qsp-qsvt-h3}.

This comparison should be interpreted carefully. Figure~\ref{fig:h3-cost} shows that the centered-difference Pauli-decomposition method is substantially less costly than structure-agnostic QSD for the sizes considered here. At the same time, it indicates that a different implementation route could become preferable if an efficient sparse block encoding of the friction term $\Hthree$ were available.

\subsection{\texorpdfstring{One-Step Circuit Costs for $\NVE$ and $\NVT$}{One-Step Circuit Costs for NVE and NVT}}\label{sec:nve-nvt-step-cost}

\begin{figure}[t]
  \centering
  \includegraphics[width=\linewidth]{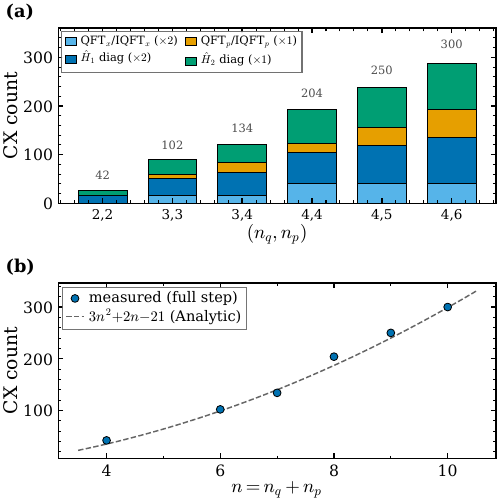}
  \caption{\textbf{CX cost of the $\NVE$ one-step propagator.}
  (a) Component-wise breakdown. The contributions from $\mathcal F_q$ and $\mathcal F_q^\dagger$, the bilinear phase of $\hat H_1$, $\mathcal F_p$ and $\mathcal F_p^\dagger$, and the diagonal phase of $\hat H_2$ are shown separately. The $n_x$ in the figure corresponds to $n_q$ in the text. The explicit gate-count benchmark uses the one-dimensional single-particle cosine potential in Eq.~\eqref{eq:cosine-cost-benchmark}. (b) Total CX count as a function of $n=n_q+n_p$. The measured values follow $3n^2+2n-21$, showing that the $\NVE$ one-step circuit grows quadratically with the number of qubits.}
  \label{fig:nve-cost}
\end{figure}

\begin{figure}[t]
  \centering
  \includegraphics[width=\linewidth]{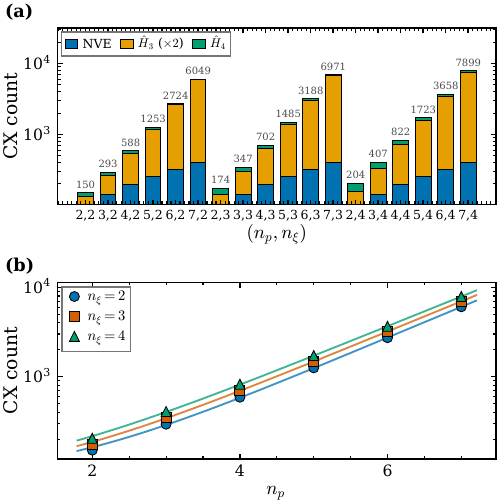}
  \caption{\textbf{CX cost of the $\NVT$ one-step propagator.}
  (a) Stacked breakdown of the CX count. The split-operator part, which has the same form as in $\NVE$, the two occurrences of the $\Hthree$ dilation block, and the thermostat block $\hat H_4$ are shown separately. The explicit gate-count benchmark uses the one-dimensional single-particle cosine potential in Eq.~\eqref{eq:cosine-cost-benchmark}. (b) Total CX count as a function of $n_p$. The series for $n_\xi=2,3,4$ follow fits of the form $n_p\,2^{n_p}$. This reflects the $n_p\,2^{n_p}$-type growth of the weighted Pauli sum in the centered-difference Pauli evolution.}
  \label{fig:nvt-cost}
\end{figure}

Here, we distinguish the asymptotic scaling implied by the circuit construction from the benchmark implementation used for the explicit gate-count measurements. The general expressions are written for $N_f$ classical degrees of freedom. By contrast, the explicit circuit syntheses in Figs.~\ref{fig:nve-cost} and~\ref{fig:nvt-cost} use the one-dimensional single-particle cosine potential:
\begin{equation}
  V(q)=V_0\cos q,
  \qquad
  F(q)=-\partial_qV(q)=V_0\sin q.
  \label{eq:cosine-cost-benchmark}
\end{equation}
The CX counts reported below are the values for the implementation of Eq.~\eqref{eq:cosine-cost-benchmark}, while the scaling with register size follows from the construction of the individual circuit blocks.

The one-step $\NVE$ propagator is $U_1(\Delta t/2)U_2(\Delta t)U_1(\Delta t/2)$ in Fig.~\ref{fig:circuits}(a). In $U_1$, the Fourier transform on the $q$ register diagonalizes $\hat\lambda_q$ as $\hat k_q$, and the phase $\Phi_q(\delta t)=\exp[-\ii\delta t\hat p\hat k_q/m]$ is applied. In $U_2$, the Fourier transform on the $p$ register maps $\hat\lambda_p$ to $\hat k_p$, and the phase $\Phi_p(\delta t)=\exp[-\ii\delta tF(\hat q)\hat k_p]$ is applied. For Eq.~\eqref{eq:cosine-cost-benchmark}, $F(\hat q)$ is implemented as a diagonal function on the $q$ register. Because these phases can be written as functions of the computational-basis values of the registers, they need not be synthesized as generic diagonal unitaries. The same structure is used in the split-operator quantum simulation of chemical dynamics by Kassal \textit{et al.} and in the diagonal-unitary synthesis of Welch \textit{et al.}~\cite{Kassal2008,Welch2014}.

Fig.~\ref{fig:nve-cost} shows the CX count of the $\NVE$ one-step circuit. The measured values are described by
\begin{equation}
  N_{\rm CX}^{\NVE}(n)=3n^2+2n-21,
  \qquad n=n_q+n_p.
  \label{eq:nve-cx}
\end{equation}
As seen from the breakdown in Fig.~\ref{fig:nve-cost}(a), the QFT/IQFT blocks and the bilinear-phase components are all $\bigO (n^2)$. Therefore, the one-step implementation of $\NVE$ remains polynomial in the number of register qubits. This means that, while the circuit represents a grid with $N_p=2^{n_p}$ momentum points, the gate count can be kept polynomial in $n_p$ rather than in $N_p$.

The one-step $\NVT$ propagator is $U_1U_2U_3U_4U_3U_2U_1$ in Fig.~\ref{fig:circuits}(b). The blocks $U_1$ and $U_2$ are the same split-operator blocks as in $\NVE$, and $U_4$ is implemented as a diagonal phase by applying a Fourier transform to the $\xi$ register,
\begin{equation}
  \hat H_4=\frac{(\hat p^2/m)-N_fT_0}{Q}\hat\lambda_\xi.
\end{equation}
These blocks have polynomial cost. The newly dominant contribution in $\NVT$ is the twice-appearing block $U_3(\Delta t/2)=\exp[-\ii(\Delta t/2)\Hthree]$.

Fig.~\ref{fig:nvt-cost} shows the cost of the $\NVT$ one-step circuit. The fits for $n_\xi=2,3,4$ are
\begin{align}
  N_{\rm CX}^{\NVT}\big|_{n_\xi=2}
  &\simeq 5.06\,n_p\,2^{n_p}+11.2\,2^{n_p}+77,
  \nonumber\\
  N_{\rm CX}^{\NVT}\big|_{n_\xi=3}
  &\simeq 4.96\,n_p\,2^{n_p}+19.2\,2^{n_p}+69,
  \label{eq:nvt-cx-fits}\\
  N_{\rm CX}^{\NVT}\big|_{n_\xi=4}
  &\simeq 4.85\,n_p\,2^{n_p}+27.2\,2^{n_p}+68.
  \nonumber
\end{align}
The leading terms are all proportional to $n_p\,2^{n_p}$. The differences among the values of $n_\xi$ appear in the constant factors originating from the Pauli-$Z$ expansion of $\hat\xi$ and in secondary terms introduced by the transpiler optimization. In general,
\begin{equation}
  N_{\rm CX}^{\NVT}=\bigO(n_\xi n_p\,2^{n_p})
  \label{eq:nvt-cost-general}
\end{equation}
and, in the range shown in the figure with fixed $n_\xi$, this can be read as $\bigO (n_p\,2^{n_p})$.


Table~\ref{tab:resource-summary} summarizes the scaling of the one-step propagators and the $\Hthree$ block. The first two rows compare the $\NVE$ and $\NVT$ one-step circuits constructed in the main text. These scalings are asymptotic estimates based on the construction of the circuit blocks, whereas the numerical CX counts in Figs.~\ref{fig:nve-cost} and~\ref{fig:nvt-cost} were measured for the benchmark implementation in Eq.~\eqref{eq:cosine-cost-benchmark}. The last three rows compare different methods for implementing the $\Hthree$ block, which is the dominant contribution in $\NVT$. Here, $n=n_q+n_p$ and $n_{\rm total}=n_p+n_\xi$.

\begin{table}[ht]
\caption{\textbf{Resource scaling of the one-step propagators and the $\Hthree$ block.}
The $\NVE$ circuit is a split-operator circuit consisting of QFTs and diagonal phase gates and can be implemented with $\bigO (n^2)$ CX gates per step. In the present $\NVT$ circuit, $U_3=e^{-\ii\delta t\Hthree}$ appears twice, and the $\bigO (n_\xi n_p\,2^{n_p})$ term from the centered-difference Pauli-decomposition method is dominant.
}
\begin{ruledtabular}
\begin{tabular}{lll}
Level & Block or method & CX scaling  \\
\hline
One-step propagator & $\NVE$ & $\bigO (n^2)$  \\
One-step propagator & $\NVT$ with Pauli $\Hthree$ & $\bigO (n^2+n_\xi n_p\,2^{n_p})$  \\
\hline
$\Hthree$ block & QSD & $\bigO (4^{n_{\rm total}})$  \\
$\Hthree$ block & Centered-difference & $\bigO (n_\xi n_p\,2^{n_p})$  \\
 & Pauli decomposition & \\
$\Hthree$ block & QSP best case & $\bigO (n_{\rm total}\,2^{n_p})$ \\
\end{tabular}
\end{ruledtabular}
\label{tab:resource-summary}
\end{table}

\section{Conclusion and Outlook}\label{sec:conclusion}

In this work, we formulated Green--Kubo transport coefficients in
classical molecular dynamics as a readout problem for a quantum algorithm
by representing Liouville evolution as unitary time evolution in the
Koopman--von Neumann (KvN) framework. Both $\NVE$ dynamics and Nos\'e--Hoover-type $\NVT$ dynamics were described by Hermitian KvN generators within a common phase-space Hilbert-space framework, with the latter acting on the extended phase space including the thermostat variable. In finite-grid numerical benchmarks, we identified time regimes in which the KvN correlation
functions agree with classical MD references, and we observed algebraic
decay of the correlation-function discretization error with the number
of grid points \(N_z\). Thus, when the grid is represented by a quantum
register with \(N_z=2^{n_z}\), this finite-grid error decreases
exponentially with the number of qubits \(n_z\) assigned to that axis.

QPE provides the readout primitive for the transport coefficient in this formulation. 
When the flux-excited state is supplied as the system-register input, the ancilla bin-zero probability \(P_0\) determines the Bartlett-windowed Green--Kubo estimator. 
For the diffusion coefficient, this relation gives
\begin{equation}
  D_{\rm Bart}
  =
  \frac{1}{2}C_{vv}(0)\tau_{\QPE}P_0.
  \label{eq:conclusion-dbart}
\end{equation}
In this way, evaluating the Green--Kubo integral is reformulated as the task of estimating the QPE-defined probability \(P_0\), rather than as a sequential integration of the time-correlation function.

Given a canonical-state-preparation oracle and the QPE oracle defining
\(P_0\), we numerically verified the quadratic query-complexity
improvement provided by MLAE for the statistical estimation of \(P_0\).
Direct shot sampling reduces the estimation error as
\(N_{\rm queries}^{-1/2}\), whereas MLAE gives scaling close to
\(N_{\rm queries}^{-1}\) in the numerical regime studied here.
Consequently, for the statistical error of the corresponding
Bartlett-windowed Green--Kubo estimator, the required number of queries
is improved from \(O(\epsilon^{-2})\) to \(O(\epsilon^{-1})\).

For $\NVT$ dynamics, we showed that classical dynamics with a
Nos\'e--Hoover thermostat can also be represented as unitary KvN
evolution. In the present centered-difference Pauli-decomposition
implementation, the friction term \(\Hthree\) has a gate cost scaling as
\(O(n_\xi n_p\,2^{n_p})\). Within the explicit circuit syntheses examined
in this work, this term is the dominant cost of the $\NVT$ one-step
propagator. Reading correlation functions directly under $\NVT$
dynamics with polynomial gate cost would therefore require a
polynomial-cost block encoding of \(\Hthree\), or an alternative
discretization or implementation of the thermostat term.

Two main directions remain for future work. The first is the efficient
construction of the canonical-state-preparation oracle. This cost must be
assessed separately from the query complexity of the Green--Kubo readout
analyzed here. The second is to connect the KvN-MD framework to
electronic-structure calculations performed on a quantum computer, by
providing a concrete interface through which forces, potential-energy
surfaces, or free-energy information are passed to the nuclear KvN
dynamics. Such an interface would allow the quantum computation of
electronic-structure information and the Green--Kubo readout formulated
in this work to be combined within a single transport-coefficient
calculation.

\begin{acknowledgments}
The author thanks Naoki Mitsumoto and Hidehiko Hiramatsu of DENSO CORPORATION, as well as the members of Quemix Inc., for valuable contributions and discussions. This work was supported by the Center of Innovation for Sustainable Quantum AI, JST Grant Number JPMJPF2221. The computation was performed using the facilities of the Supercomputer Center, the Institute for Solid State Physics, the University of Tokyo (ISSPkyodo-SC-2026-Ea-0014), the TSUBAME4.0 supercomputer at the Institute of Science Tokyo, and the Supermicro ARS-111GL-DNHR-LCC and FUJITSU Server PRIMERGY CX2550 M7 (Miyabi) at Joint Center for Advanced High Performance Computing (JCAHPC).
\end{acknowledgments}
\appendix
\section{Stability of Discretized KvN Dynamics}
\label{app:discretization-stability}
\label{sec:discretization-stability}

In this appendix, we examine the stability of the finite-grid KvN
evolution used in the main text. On a quantum computer, phase space is
represented by a finite set of grid points, and each phase-space axis is
encoded in a finite qubit register. It is therefore useful to separate
the errors arising from boundary truncation, grid resolution, and time
discretization. We use a one-particle model to identify which numerical
parameters control these errors.

We mainly consider three types of errors. The first is the wrap-around
error caused by truncating nonperiodic axes, such as the momentum \(p\)
and the thermostat variable \(\xi\), to finite intervals. The second is
the grid-resolution error associated with representing the wave function
on a finite number of grid points. The third is the time-discretization
error introduced when the continuous-time evolution
\(\exp(-\ii \Hkvn t)\) is approximated by the second-order symmetric
Suzuki--Trotter product formula. Our purpose is not to combine these
contributions into a single error estimate, but rather to distinguish
their physical origins and the parameters that suppress them.

\subsection{Test Model and Error Metric}\label{sec:discretization-model}

For the test system, we use the one-dimensional cosine potential
\begin{equation}
  V(q)=V_0\cos q,
  \qquad q\in[0,2\pi).
  \label{eq:cosine-model}
\end{equation}
The coordinate \(q\) is a physically periodic variable. By contrast, the
momentum \(p\) and the Nos\'e--Hoover variable \(\xi\) are nonperiodic
variables and are numerically truncated to the finite intervals
\begin{equation}
  p\in[-p_{\max},p_{\max}),
  \qquad
  \xi\in[-\xi_{\max},\xi_{\max}).
  \label{eq:finite-nonperiodic-domain}
\end{equation}
We use \(V_0=m=T_0=Q=1\) as the reference parameter set. The thermal
widths are
\begin{equation}
  \sigma_p=\sqrt{mT_0},
  \qquad
  \sigma_\xi=\sqrt{T_0/Q},
  \label{eq:thermal-widths}
\end{equation}
and the boundary widths are specified as
\(p_{\max}=n_\sigma\sigma_p\) and
\(\xi_{\max}=n_\sigma\sigma_\xi\).

The initial state is the equilibrium state, evaluated on the grid, that
would be returned by the canonical-state-preparation oracle assumed in
Sec.~\ref{sec:state-preparation-oracle}. Thus, for \(\NVE\), we use
\(\psi_{\rm can}(q,p)\propto\exp[-\beta H_{\rm cl}(q,p)/2]\), whereas for
\(\NVT\), we use the corresponding extended state obtained by multiplying
this state by \(\exp[-\beta Q\xi^2/4]\). The \(\NVT\) calculations in this
appendix are intended to test the stability of time evolution generated by
the Nos\'e--Hoover KvN generator on a discrete grid. They do not assume an
efficient procedure for preparing the canonical state from an arbitrary
initial state.

As a stability metric, we use the kinetic temperature
\begin{equation}
  T_{\rm kin}(t)
  =\frac{1}{N_f}\sum_{i=1}^{N_f}
   \left\langle \frac{\hat p_i^2}{m_i}\right\rangle_t .
  \label{eq:kinetic-temperature}
\end{equation}
Since \(T_{\rm kin}=T_0\) in canonical equilibrium, we use the relative
drift at a fixed observation time \(T_{\rm sim}\),
\begin{equation}
  \delta_T
  =\frac{|T_{\rm kin}(T_{\rm sim})-T_0|}{T_0},
  \label{eq:temperature-drift}
\end{equation}
as a representative measure of the discretization error. In
Fig.~\ref{fig:convergence}, we set \(T_{\rm sim}=240\) and compare three
time steps, \(\Delta t=0.01,0.05,0.1\).

\subsection{Boundaries of Nonperiodic Axes and Wrap-Around Error}\label{sec:wraparound-error}

The \(q\) axis is physically periodic and is therefore compatible with
Fourier differentiation. By contrast, the \(p\) and \(\xi\) axes are
intrinsically nonperiodic. When Fourier differentiation is used on a
finite grid, however, these finite intervals are numerically treated as
periodically connected. Consequently, if the tail of the probability
density reaches a boundary, an unphysical wrap-around from \(+z_{\max}\)
to \(-z_{\max}\) occurs, where \(z\) denotes either \(p\) or \(\xi\).

This effect is particularly transparent for the Nos\'e--Hoover variable.
The region \(\xi>0\) is a cooling region in which momenta are damped,
whereas \(\xi<0\) is a heating region in which momenta are amplified. If
the tail of the \(\xi\) distribution wraps from \(+\xi_{\max}\) to
\(-\xi_{\max}\), components on the cooling side are unphysically
transferred to the heating side, producing a temperature drift. Similarly,
wrap-around along the \(p\) axis distorts high-momentum components and
their correlation with the force field, thereby degrading long-time
temperature stability.

The nonperiodic components of the canonical distribution have Gaussian
tails, so the relative density at the boundary is suppressed as
\begin{equation}
  \frac{\varrho(z_{\max})}{\varrho(0)}
  =\exp\left(-\frac{n_\sigma^2}{2}\right).
  \label{eq:gaussian-boundary-tail}
\end{equation}
Thus, as a first approximation, the temperature drift caused by the
boundary decreases as
\begin{equation}
  \delta_T^{\rm wrap}
  \simeq A_z\exp\left(-\frac{n_\sigma^2}{2}\right).
  \label{eq:wraparound-scaling}
\end{equation}
Here \(A_z\) is a coefficient that depends on the force field, the
observation time, and the time step. This exponential dependence is not a
strict universal law, but it is useful as a diagnostic when the dominant
error source is the probability mass reaching the boundary.

\begin{figure}[htbp]
  \centering
  \includegraphics[width=\linewidth]{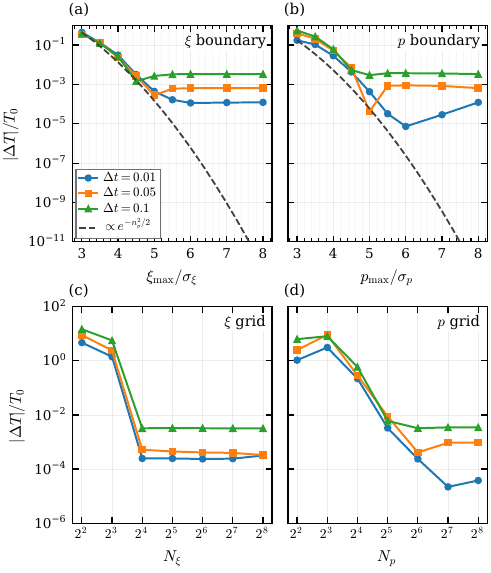}
  \caption{\textbf{Stability of discretized KvN dynamics.}
  The test system is a one-particle \(\NVT\) cosine potential with
  \(V_0=m=T_0=Q=1\), \(T_{\rm sim}=240\), and
  \(\Delta t\in\{0.01,0.05,0.1\}\).
  (a) Boundary-width scan for the \(\xi\) axis.
  (b) Boundary-width scan for the \(p\) axis.
  The dashed line is the Gaussian-tail diagnostic proportional to
  \(\exp(-n_\sigma^2/2)\).
  (c) Grid-size scan for the \(\xi\) axis.
  (d) Grid-size scan for the \(p\) axis.
  For sufficiently large boundary widths and grid sizes, the temperature
  drift reaches a floor determined by \(\Delta t\).}
  \label{fig:convergence}
\end{figure}

Figures~\ref{fig:convergence}(a) and \ref{fig:convergence}(b) directly
test this estimate by scanning the boundary width. For both the \(\xi\)
and \(p\) axes, the temperature drift decreases approximately along
\(\exp(-n_\sigma^2/2)\) in the range \(n_\sigma=3\)--\(5\), and for
\(n_\sigma\gtrsim6\), the boundary error becomes smaller than the other
error sources. In the calculations below, we therefore use
\begin{equation}
  p_{\max}\gtrsim 6\sigma_p,
  \qquad
  \xi_{\max}\gtrsim 6\sigma_\xi
  \label{eq:six-sigma-rule}
\end{equation}
as reference boundary conditions for the nonperiodic axes. These values
are also consistent with empirical velocity-space truncation rules used
in Fourier-type continuum Vlasov calculations~\cite{Cheng1976,Klimas1987,Klimas1994,Vogman2014}.

\subsection{Grid Resolution and the Time-Discretization Floor}\label{sec:grid-and-trotter-error}

After the boundary width is made sufficiently large, the remaining errors
are controlled by the number of grid points and the time step.
Figures~\ref{fig:convergence}(c) and \ref{fig:convergence}(d) show the
results obtained by fixing either \(\xi_{\max}\) or \(p_{\max}\) to
\(8\sigma\) and varying \(N_\xi\) and \(N_p\). In the low-resolution
regime, even when the Gaussian tail is contained within the truncated
interval, the wave function and its derivatives are not sufficiently
resolved, and the temperature drift remains large. In contrast, the drift
decreases rapidly for \(N_\xi,N_p\gtrsim 16\), and for
\(N_\xi,N_p\gtrsim 32\), it nearly reaches the floor associated with each
value of \(\Delta t\).

This floor is consistent with the time-discretization error of the
second-order symmetric Suzuki--Trotter product formula. The
\(S_2(\Delta t)\) used in Sec.~\ref{sec:dilation-and-nvt-product} has a
local error of \(\bigO(\Delta t^3)\) per step, and therefore the global
error over a fixed physical time \(T\) is expected to scale as
\begin{equation}
  \delta_T^{\rm ST}
  \sim C_{\rm ST}\,T\Delta t^2+\bigO(\Delta t^4).
  \label{eq:st-error-scaling}
\end{equation}
\cite{Trotter1959,Suzuki1976,Suzuki1990,Yoshida1990,Kivlichan2019}.
In each panel of Fig.~\ref{fig:convergence}, the floor reached after the
boundary is made sufficiently wide or the grid sufficiently fine increases
in the order \(\Delta t=0.01,0.05,0.1\), supporting this interpretation.

In summary, the temperature drift near equilibrium can be practically
decomposed as
\begin{align}
  \delta_T
  \simeq{}&
  C_{\rm ST}T\Delta t^2
  +A_p\exp\left(-\frac{p_{\max}^2}{2\sigma_p^2}\right) \nonumber\\
  &+A_\xi\exp\left(-\frac{\xi_{\max}^2}{2\sigma_\xi^2}\right)
  +\delta_T^{\rm grid}.
  \label{eq:discretization-error-summary}
\end{align}
Here \(\delta_T^{\rm grid}\) is the residual contribution from finite grid
resolution. In the settings of Figs.~\ref{fig:convergence}(c) and
\ref{fig:convergence}(d), it becomes smaller than the time-discretization
floor for \(N_p,N_\xi\gtrsim 32\). Although this expression is not a
rigorous error bound, it serves as a practical design formula for tuning
the boundary width, grid resolution, and time step independently.

\subsection{Recommended Discretization and the \texorpdfstring{$N=2^n$}{N = 2 to the n} Grid Representation}\label{sec:recommended-grid}

Table~\ref{tab:discretization-guideline} summarizes the practical
discretization conditions inferred from Fig.~\ref{fig:convergence} for
the one-particle one-dimensional cosine potential. These values are
representative for the present model. Systems with stiffer potentials,
lower-temperature distributions, or longer correlation times require
separate convergence checks for the boundary width, number of grid points,
and time step.

\begin{table}[t]
  \caption{\textbf{Guidelines for the discretization conditions in the
  one-particle one-dimensional cosine potential.}
  The boundary widths of the nonperiodic axes are expressed in terms of
  the thermal widths \(\sigma_p=\sqrt{mT_0}\) and
  \(\sigma_\xi=\sqrt{T_0/Q}\).}
  \label{tab:discretization-guideline}
  \begin{ruledtabular}
  \begin{tabular}{@{}lll@{}}
    Quantity & Guideline & Control \\
    \hline
    \(p_{\max}\) & \(\gtrsim 6\sigma_p\) & \(p\) wrap-around \\
    \(\xi_{\max}\) & \(\gtrsim 6\sigma_\xi\) & \(\xi\) wrap-around \\
    \(N_p\) & \(\gtrsim 32\) & \(p\) resolution \\
    \(N_\xi\) & \(\gtrsim 32\) & \(\xi\) resolution \\
    \(N_q\) & Resolve modes of \(V(q)\) & \(q\) resolution \\
    \(\Delta t\) & Converge \(S_2(\Delta t)\) & time-discretization floor
  \end{tabular}
  \end{ruledtabular}
\end{table}

The important point for quantum computation is that the number of grid
points along each axis grows with the number of qubits as
\begin{equation}
  N_z=2^{n_z}.
  \label{eq:grid-qubit-relation}
\end{equation}
Thus, increasing the grid resolution from \(N_z\) to \(2N_z\) requires
adding only one qubit to the corresponding register. In a classical
state-vector representation, memory grows in proportion to the product of
the numbers of grid points as the phase-space dimension increases. In a
quantum state representation, by contrast, the grid is encoded in qubit
registers. This distinction becomes important when large \(N_z\) is
needed to suppress discretization errors.

This statement does not mean that discretization errors disappear
automatically. Sufficient boundary widths, sufficient grid resolution, and
a sufficiently small \(\Delta t\) are still required. Moreover, even if
the grid can be encoded using \(n_z\) qubits, a quantum-computational
advantage can be realized only if the time-evolution operator is
implemented with polynomial gate cost. As shown in the resource analysis,
the \(\NVE\) one-step propagator admits an \(\bigO(n^2)\) split-operator
implementation, whereas in \(\NVT\), the term
\(\hat H_3=-\hat\xi\otimes\hat D_p\) becomes an
\(\bigO(n_\xi n_p\,2^{n_p})\) bottleneck in the present implementation.
Thus, the conclusion of this appendix is twofold: it gives grid
conditions under which the discretization error can be controlled, and it
separates the representation-level benefit of quantum registers from the
remaining implementation bottleneck in the \(\NVT\) propagator.

In the actual diffusion-coefficient readout in this paper, we assume as
input the equilibrium state supplied by the canonical-state-preparation
oracle, and we mainly use the \(\NVE\) propagator for the subsequent
correlation-function evaluation. Therefore, the \(\NVT\) stability
analysis is intended to validate the Nos\'e--Hoover KvN generator itself;
it does not imply that the controlled unitary in QPE must always be the
\(\NVT\) propagator. In Sec.~\ref{sec:dynamic-correlations}, we test,
under these discretization conditions, how accurately the velocity
autocorrelation function is reproduced as a KvN inner product.


\section{Derivation of the QPE Bin-Zero Probability and the Bartlett Window}\label{app:qpe-bartlett}

In the main text, we used the fact that the bin-zero probability obtained
by applying QPE to the flux-excited state \(\ket{\alpha_{\cal D}}\)
gives a finite-window Bartlett-windowed Green--Kubo estimator. This
appendix gives the algebraic derivation of
Eqs.~\eqref{eq:dirichlet-square-identity}--\eqref{eq:D-bartlett-trapezoid}
in the main text.

We first define the finite Dirichlet kernel
\begin{equation}
  D_K(\theta)
  =
  \sum_{r=0}^{K-1}e^{-\ii r\theta}.
\end{equation}
Using the finite geometric series, this can be written as
\begin{equation}
  D_K(\theta)
  =
  e^{-\ii (K-1)\theta/2}
  \frac{\sin(K\theta/2)}{\sin(\theta/2)}.
\end{equation}
Therefore,
\begin{equation}
  |D_K(\theta)|^2
  =
  \frac{\sin^2(K\theta/2)}{\sin^2(\theta/2)}.
  \label{eq:app-dirichlet-ratio}
\end{equation}
For \(\theta=2\pi n\), with integer \(n\), this expression is understood
by continuity as \(|D_K|^2=K^2\).

The same quantity can also be expressed as a double sum:
\begin{align}
  |D_K(\theta)|^2
  &=
  \sum_{r=0}^{K-1}\sum_{r'=0}^{K-1}
  e^{-\ii(r-r')\theta}.
\end{align}
For a fixed difference \(s=r-r'\), the number of pairs \((r,r')\) with
that difference is \(K-|s|\), for \(|s|\le K-1\). Hence,
\begin{align}
  |D_K(\theta)|^2
  &=
  \sum_{s=-(K-1)}^{K-1}
  (K-|s|)e^{-\ii s\theta}  \\
  &=
  K+
  \sum_{s=1}^{K-1}
  (K-s)
  \left(e^{-\ii s\theta}+e^{+\ii s\theta}\right) \\
  &=
  K+2\sum_{s=1}^{K-1}(K-s)\cos(s\theta).
  \label{eq:app-dirichlet-convolution}
\end{align}
Equations~\eqref{eq:app-dirichlet-ratio} and
\eqref{eq:app-dirichlet-convolution} are the squared-Dirichlet-kernel
identities used in the main text.

Next, let
\(U_{\cal D}\ket{\phi_k}=e^{-\ii\theta_k}\ket{\phi_k}\), and expand the
flux-excited state as
\begin{equation}
  \ket{\alpha_{\cal D}}
  =
  \sum_k a_k\ket{\phi_k}.
\end{equation}
For an \(m_{\rm anc}\)-qubit QPE circuit with
\(K=2^{m_{\rm anc}}\), the bin-zero probability is
\begin{equation}
  P_0
  =
  \sum_k |a_k|^2
  \frac{|D_K(\theta_k)|^2}{K^2}.
\end{equation}
Substituting Eq.~\eqref{eq:app-dirichlet-convolution} gives
\begin{align}
  P_0
  &=
  \frac{1}{K^2}
  \left[
    K+
    2\sum_{s=1}^{K-1}(K-s)
    \sum_k |a_k|^2\cos(s\theta_k)
  \right].
\end{align}
On the other hand,
\begin{align}
  \mathrm{Re}\,
  \bra{\alpha_{\cal D}}U_{\cal D}^{s}\ket{\alpha_{\cal D}}
  &=
  \mathrm{Re}
  \sum_k |a_k|^2 e^{-\ii s\theta_k} \\
  &=
  \sum_k |a_k|^2\cos(s\theta_k)
  =
  c_{\cal D}(s\Delta t).
\end{align}
Thus, the bin-zero probability can be written as
\begin{equation}
  P_0
  =
  \frac{1}{K^2}
  \left[
    K+2\sum_{s=1}^{K-1}(K-s)c_{\cal D}(s\Delta t)
  \right].
  \label{eq:app-p0-bartlett}
\end{equation}

Finally, multiplying Eq.~\eqref{eq:app-p0-bartlett} by
\(\frac{1}{2}C_{JJ}^{({\cal D})}(0)\tau_{\QPE}\) and using
\(\tau_{\QPE}=K\Delta t\), we obtain
\begin{align}
  \frac{1}{2}C_{JJ}^{({\cal D})}(0)\tau_{\QPE}P_0
  &=
  C_{JJ}^{({\cal D})}(0)\Delta t \nonumber\\
  &\quad\times
  \left[
    \frac{1}{2}
    +\sum_{s=1}^{K-1}
    \left(1-\frac{s}{K}\right)c_{\cal D}(s\Delta t)
  \right].
\end{align}
This is the discrete Green--Kubo estimator obtained by applying the
Bartlett window
\[
  w_s=1-\frac{s}{K},
  \qquad s=0,1,\ldots,K,
\]
together with the trapezoidal-rule coefficient \(1/2\) at \(s=0\).
The window vanishes at \(s=K\), so no endpoint contribution appears there.
For a diffusion coefficient, setting \(C_{JJ}=C_{vv}\) gives
Eq.~\eqref{eq:qpe-bin0-main-formula} in the main text.

\section{Details of the Centered-Difference Pauli-Decomposition Method}\label{app:central-difference-pauli}

This appendix summarizes the centered-difference Pauli-decomposition method used in Sec.~\ref{sec:h3-pauli}. The target operator is the momentum-dilation generator appearing in the Nos\'e--Hoover friction term,
\begin{equation}
  \Hthree=-\hat\xi\otimes \hat D_p,
  \qquad
  \hat D_p=\frac{1}{2}\left(\hat p\hat\lambda_p+\hat\lambda_p\hat p\right).
\end{equation}
Here \(n_p\) denotes the number of qubits in the momentum register, and \(N_p=2^{n_p}\) is the number of momentum grid points.

In the benchmark code used for the resource analysis, the momentum grid is defined as
\begin{align}
  p_j &= p_{\min}+j\Delta p, \nonumber\\
  \Delta p &= \frac{p_{\max}-p_{\min}}{N_p},
  \qquad j=0,\ldots,N_p-1,
  \label{eq:app-p-grid}
\end{align}
and a cyclic centered difference is used as the finite-difference operator. That is, the indices are treated modulo \(N_p\), so a wrap-around bond between \(j=N_p-1\) and \(j=0\) is included. This is the same convention as that used in the circuit-resource measurements in Fig.~\ref{fig:h3-cost}.

With this convention, the Hermitian discrete dilation operator is
\begin{align}
  \hat D_p^{\rm cyc}
  &=
  -\ii\sum_{j=0}^{N_p-1}
  \gamma_j
  \left(
    \ket{j+1}\bra{j}-\ket{j}\bra{j+1}
  \right), \nonumber\\
  \gamma_j&=\frac{p_j+p_{j+1}}{4\Delta p},
  \label{eq:app-Dp-cyc}
\end{align}
where \(j+1\) is understood modulo \(N_p\). Equation~\eqref{eq:app-Dp-cyc} is the centered-difference discretization of the continuous operator \((\hat p\hat\lambda_p+\hat\lambda_p\hat p)/2\). If a nonperiodic momentum window is used, the endpoint treatment can be modified separately. The gate counts in Figs.~\ref{fig:h3-cost} and~\ref{fig:nvt-cost}, however, are based on the cyclic convention in Eq.~\eqref{eq:app-Dp-cyc}.

We expand Eq.~\eqref{eq:app-Dp-cyc} in the Pauli basis on the \(n_p\)-qubit computational basis:
\begin{equation}
  \hat D_p^{\rm cyc}
  =\sum_{\mu\in{\cal S}_{D_p}} a_\mu P_\mu,
  \qquad
  P_\mu\in\{I,X,Y,Z\}^{\otimes n_p},
  \quad
  a_\mu\in\mathbb{R}.
  \label{eq:app-Dp-pauli}
\end{equation}
Equation~\eqref{eq:app-Dp-cyc} is a purely imaginary antisymmetric matrix. Therefore, only Pauli strings containing an odd number of \(Y\) operators can have nonzero coefficients. Indeed, if \(N_Y(P_\mu)\) denotes the number of \(Y\) operators in \(P_\mu\), then
\[
  P_\mu^*=(-1)^{N_Y(P_\mu)}P_\mu .
\]
Pauli strings with an even number of \(Y\) operators are real matrices and have zero Hilbert--Schmidt inner product with Eq.~\eqref{eq:app-Dp-cyc}. We refer to this selection rule as the odd-\(Y\) filter.

The number of terms in the Pauli expansion can be counted by classifying the nearest-neighbor bonds \(\ket{j+1}\bra{j}\) according to the carry length in the binary representation of \(j\). A longer carry in the update \(j\to j+1\) produces Pauli strings with larger weight. The cyclic bond \(N_p-1\to0\) introduces the longest carry and therefore increases the number of Pauli terms compared with a nonperiodic finite-difference operator. Combining this carry structure with the odd-\(Y\) filter gives the number of nonzero Pauli strings as
\begin{equation}
  N_{D_p}^{\rm cyc}(n_p)
  \equiv |{\cal S}_{D_p}|
  =\frac{7}{4}2^{n_p}-(n_p+2),
  \qquad n_p\ge 2.
  \label{eq:app-N-Dp}
\end{equation}
This closed form corresponds to the Pauli decomposition used in the Pauli-evolution benchmark in Fig.~\ref{fig:h3-cost}.

The circuit-cost estimate depends not only on the number of Pauli terms but also on the Pauli-string weights. We define the weight \(w(P_\mu)\) as the number of nonidentity Pauli operators in \(P_\mu\), and define the weighted Pauli sum by
\begin{equation}
  W_{D_p}^{\rm cyc}(n_p)
  \equiv
  \sum_{\mu\in{\cal S}_{D_p}} w(P_\mu).
  \label{eq:app-W-Dp-definition}
\end{equation}
A single Pauli rotation \(\exp(-\ii\theta P_\mu)\) can be implemented using basis changes, a CNOT ladder, one \(R_z\) rotation, and the inverse ladder. Thus, \(w(P_\mu)\) determines the leading contribution to the two-qubit gate count for that rotation.

For the Pauli expansion of Eq.~\eqref{eq:app-Dp-cyc}, the weighted Pauli sum is
\begin{equation}
  W_{D_p}^{\rm cyc}(n_p)
  =\frac{7}{4}n_p\,2^{n_p}-3\cdot 2^{n_p}+3,
  \qquad n_p\ge 2.
  \label{eq:app-W-Dp}
\end{equation}
Therefore, the centered-difference Pauli-decomposition method gives
\begin{equation}
  N_{D_p}^{\rm cyc}(n_p)=\bigO(2^{n_p}),
  \qquad
  W_{D_p}^{\rm cyc}(n_p)=\bigO(n_p\,2^{n_p}).
\end{equation}
Equivalently, since \(N_p=2^{n_p}\), the weighted Pauli sum scales as
\[
  \bigO(N_p\log N_p)
\]
with the number of momentum grid points, and hence exponentially with the number of momentum qubits \(n_p\).

Finally, we include the tensor product with \(\hat\xi\). The coordinate operator \(\hat\xi\) is diagonal on the \(\xi\) register, and under binary encoding it can be expanded as
\begin{equation}
  \hat\xi=
  \sum_{\nu\in{\cal S}_\xi} b_\nu Z_\nu,
  \qquad
  Z_\nu\in\{I,Z\}^{\otimes n_\xi}.
  \label{eq:app-xi-pauli}
\end{equation}
For a coordinate operator on an equally spaced grid, the number of nonzero \(Z\)-string terms is \(\bigO(n_\xi)\). Hence
\begin{equation}
  \Hthree
  =-
  \sum_{\nu\in{\cal S}_\xi}
  \sum_{\mu\in{\cal S}_{D_p}}
  b_\nu a_\mu\, Z_\nu\otimes P_\mu,
  \label{eq:app-H3-pauli}
\end{equation}
and the leading cost of a Pauli-evolution implementation is
\begin{equation}
  \bigO\!\left(n_\xi W_{D_p}^{\rm cyc}(n_p)\right)
  =
  \bigO\!\left(n_\xi n_p\,2^{n_p}\right).
  \label{eq:app-H3-cost}
\end{equation}
The growth of the \(\NVT\) cost shown in Figs.~\ref{fig:h3-cost} and~\ref{fig:nvt-cost} originates from this exponential dependence on the momentum-register size \(n_p\) in the present centered-difference Pauli-decomposition implementation.

\section{Evaluation Method and Oracle Assumptions for the QSP Best-Case Line}\label{app:qsp-qsvt-h3}

This appendix summarizes how the QSP best-case line in
Fig.~\ref{fig:h3-cost} is evaluated. The green series in that figure is
not obtained by transpiling an implemented quantum circuit in Qiskit.
Rather, it is a theoretical CX estimate for Hamiltonian simulation by QSP
under a sparse-access oracle model for the sparse-matrix representation
of \(\Hthree\).

The target operator is
\begin{equation}
  \Hthree=-\hat\xi\otimes\hat D_p.
  \label{eq:app-qsp-H3}
\end{equation}
The \(\hat D_p\) used in the benchmark code is the nearest-neighbor
finite-difference matrix based on the cyclic centered difference, with
nonzero bond coefficients
\begin{equation}
  \gamma_j=\frac{p_j+p_{j+1}}{4\Delta p},
  \qquad j+1\equiv j+1\pmod{N_p}.
  \label{eq:app-qsp-gamma}
\end{equation}
The row sparsity with respect to the \(p\) register is therefore \(d=2\).

For the best-case line, we assume the following two sparse-access oracles:
\begin{align}
  O_{\rm loc}:
  \ket{i}\ket{\ell}
  &\mapsto
  \ket{i}\ket{j_\ell(i)},
  \qquad \ell=0,1,
  \label{eq:app-qsp-Oloc}\\
  O_{\rm val}:
  \ket{i}\ket{j}\ket{0}
  &\mapsto
  \ket{i}\ket{j}\ket{(\Hthree)_{ij}}.
  \label{eq:app-qsp-Oval}
\end{align}
Here \(O_{\rm loc}\) returns the nonzero column indices in row \(i\), and
\(O_{\rm val}\) returns the corresponding matrix element to fixed-point
precision. This oracle model differs from a Pauli-LCU circuit that
directly implements a sum of Pauli strings through SELECT operations.
Thus, the best-case line in Fig.~\ref{fig:h3-cost} should be regarded as
a theoretical target that would be reachable only if these sparse-access
oracles were available at low cost.

In the sparse-access model, we use an upper bound based on the row-sum
norm as the block-encoding normalization factor. In the benchmark of this
work, this factor is taken as
\begin{equation}
  \alpha_{\rm sparse}
  =2\|\Hthree\|_{\max}
  =2\,\xi_{\max}\max_j|\gamma_j|.
  \label{eq:app-qsp-alpha-sparse-exact}
\end{equation}
Here \(\xi_{\max}=(2^{n_\xi}-1)\Delta\xi/2\). If the \(p\) and \(\xi\)
windows are fixed, then \(\Delta p=\bigO(2^{-n_p})\), and
Eq.~\eqref{eq:app-qsp-alpha-sparse-exact} gives
\(\alpha_{\rm sparse}=\bigO(2^{n_p})\).

In QSP, the normalized operator \(M=\Hthree/\alpha_{\rm sparse}\) is used
to approximate
\begin{equation}
  e^{-\ii\delta t\Hthree}
  =e^{-\ii\tau M},
  \qquad
  \tau=\delta t\,\alpha_{\rm sparse}.
  \label{eq:app-qsp-tau-sparse}
\end{equation}
In the benchmark, we set \(\delta t=\texttt{dt\_half}=0.01\), and choose
the truncation degree \(K_{\rm QSP}\) of the Jacobi--Anger expansion from
the condition
\begin{equation}
  |J_{K_{\rm QSP}+1}(\tau)|<\epsilon_{\rm qsp}/2.
  \label{eq:app-qsp-K-condition}
\end{equation}
Here \(J_m\) is the Bessel function. Asymptotically, this condition gives
\begin{equation}
  K_{\rm QSP}
  =
  \bigO\!\left(\tau+\log\frac{1}{\epsilon_{\rm qsp}}\right)
  =
  \bigO\!\left(2^{n_p}+\log\frac{1}{\epsilon_{\rm qsp}}\right).
  \label{eq:app-qsp-K-scaling}
\end{equation}
In the present setting, we use \(\epsilon_{\rm qsp}=10^{-6}\).

Following the convention of the benchmark code, the CX count for a single
sparse block-encoding primitive is estimated as
\begin{equation}
  C_{\rm BE}^{\rm sparse}
  =6n_{\rm total}+3n_{\rm total}
  +\left\lfloor
     n_{\rm total}\log_2\!\left(\frac{1}{\epsilon_{\rm qsp}}\right)
   \right\rfloor
  +n_{\rm total}.
  \label{eq:app-qsp-CBE-sparse}
\end{equation}
The four terms represent, respectively, the row oracle, the value oracle,
an RUS-arcsin-type amplitude transformation, and overheads such as
reflections. The integer rounding corresponds to the \texttt{int}
operation in the benchmark code. For \(\epsilon_{\rm qsp}=10^{-6}\), we
have \(\log_2(1/\epsilon_{\rm qsp})\simeq19.9\), so
\(C_{\rm BE}^{\rm sparse}\) is approximately \(30n_{\rm total}\).

For the full QSP estimate, we use the convention
\begin{equation}
  N_{\rm BE}^{\rm best}=4K_{\rm QSP}+3,
  \qquad
  N_{\rm refl}^{\rm best}=2K_{\rm QSP}+2.
  \label{eq:app-qsp-best-calls}
\end{equation}
Counting each reflection layer as \(4n_{\rm total}\) CX gates, the green
line in Fig.~\ref{fig:h3-cost} is given by
\begin{equation}
  N_{\rm CX}^{\rm QSP,best}
  =
  (4K_{\rm QSP}+3)C_{\rm BE}^{\rm sparse}
  +(2K_{\rm QSP}+2)4n_{\rm total}.
  \label{eq:app-qsp-best-cost}
\end{equation}
Equations~\eqref{eq:app-qsp-K-scaling} and
\eqref{eq:app-qsp-CBE-sparse} imply the leading scaling
\begin{equation}
  N_{\rm CX}^{\rm QSP,best}
  =\bigO(n_{\rm total}\,2^{n_p}).
  \label{eq:app-qsp-best-scaling}
\end{equation}

As a numerical reproduction example, take \(n_p=10\), \(n_\xi=4\),
\(\epsilon_{\rm qsp}=10^{-6}\), and \(\delta t=0.01\). Then
\(n_{\rm total}=14\), \(\alpha_{\rm sparse}=1918.125\),
\(\tau=19.18125\), and \(K_{\rm QSP}=34\). The block-encoding cost is
\begin{equation}
  C_{\rm BE}^{\rm sparse}
  =6(14)+3(14)+\lfloor 14\log_2(10^6)\rfloor+14
  =419,
  \label{eq:app-qsp-CBE-example}
\end{equation}
and Eq.~\eqref{eq:app-qsp-best-cost} gives
\begin{equation}
  N_{\rm CX}^{\rm QSP,best}
  =(4\cdot34+3)419+(2\cdot34+2)4\cdot14
  =62161.
  \label{eq:app-qsp-example}
\end{equation}
This is the same convention as that used for the best-case series in
Fig.~\ref{fig:h3-cost}.

This estimate is a theoretical reference for the case in which the sparse
structure of \(\Hthree\) can be accessed through low-cost oracles.
Therefore, in Fig.~\ref{fig:h3-cost}, it should not be interpreted as the
same type of data as the transpiled CX counts for the native Pauli circuit
or the QSD circuit. Instead, it is used as a best-case reference for a
possible implementation based on a more efficient block encoding of the
Nos\'e--Hoover friction term.

\section{Validation by Quantum-Circuit Simulation}\label{app:circuit-validation}

This appendix verifies that the circuits constructed in
Sec.~\ref{sec:circuit-implementation} reproduce the same propagators and
QPE ancilla-bin distributions as the discrete KvN state-vector
calculation. The main text presents the circuit construction; here we
collect the numerical checks of the circuit-level implementation.

\subsection{Validation Design}\label{sec:circuit-simulation-design}

The quantum-circuit simulations are performed in two stages. First, to
validate \(S_2^{\mathcal D}(\Delta t)\) without QPE ancillas, we compare
the discrete KvN state-vector calculation with the Qiskit state-vector
simulation. As the observable, we use the reduced phase-space distribution
of particle 1,
\begin{equation}
  P^{(1)}(q_1,p_1;t)
  =\sum_{q_2,p_2}\left|\braket{q_1,p_1,q_2,p_2}{\psi(t)}\right|^2.
  \label{eq:reduced-phase-space-distribution}
\end{equation}
This provides a direct comparison of the probability distribution on the
system register, independently of any QPE ancilla measurement.

Second, we validate the full QPE circuit. We compare the ancilla-bin
probability distribution \(\{P_\ell^{\rm KvN}\}\) computed from the
discrete KvN state vector with \(\{P_\ell^{\rm circ}\}\) obtained from the
Qiskit state-vector simulation. In addition, in shot-based simulations,
we estimate the bin-zero frequency \(\hat P_0\) from a finite number of
shots and examine the mean and standard deviation of
\begin{equation}
  \hat D
  =\frac{1}{2}\Cvv(0)\tau_{\QPE}\hat P_0.
  \label{eq:Dhat-from-shot-P0}
\end{equation}
The fluctuations observed in this test are the statistical errors of
Bernoulli trials with success probability \(P_0\). These are the errors
targeted by the amplitude-estimation procedure discussed in
Sec.~\ref{sec:amplitude-estimation}.

\subsection{Agreement of Reduced Phase-Space Distributions}\label{sec:phase-space-circuit}

\begin{figure}[htbp]
  \centering
  \includegraphics[width=\linewidth]{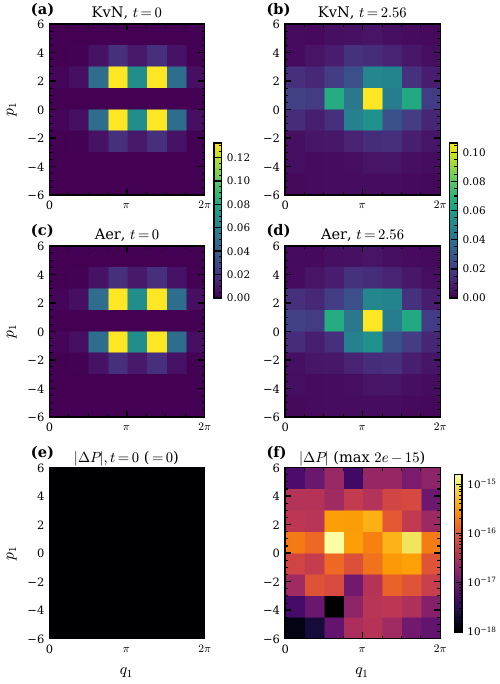}
  \caption{\textbf{Validation of the one-step propagator using reduced phase-space distributions.}
  For the two-particle coupled cosine system, we compare the reduced
  phase-space distribution \(P^{(1)}(q_1,p_1;t)\) of particle 1. The top
  row shows the discrete KvN state-vector calculation, the middle row
  shows the Qiskit state-vector simulation, and the bottom row shows the
  bin-wise differences. The difference is numerically zero at the initial
  time, and after propagation to \(t=2.56\), the maximum difference remains
  approximately \(2\times 10^{-15}\), as indicated in the figure.}
  \label{fig:phase-space-circuit}
\end{figure}

Figure~\ref{fig:phase-space-circuit} compares the reduced phase-space
distributions for the two-particle coupled cosine system. For the system
register composed of the \(q\) and \(p\) registers of each particle, we
repeatedly apply the same one-step propagator and evaluate
\(P^{(1)}(q_1,p_1;t)\) at \(t=2.56\). The distributions obtained from the
discrete KvN state-vector calculation and from the Qiskit state-vector
simulation agree after propagation, with a maximum difference of
\(\sim 2\times 10^{-15}\). This difference is within double-precision
roundoff error, confirming that the split-operator circuit reproduces the
discrete KvN propagator.

This check is independent of the QPE ancilla distribution. The QPE bin
probabilities are projections onto the eigenphase distribution, whereas
Fig.~\ref{fig:phase-space-circuit} compares the probability distribution
on the system register itself. Agreement at this level confirms that the
KvN propagator circuit has no structural error before it is used as the
controlled unitary in phase estimation.

\subsection{QPE Ancilla-Bin Distribution and Finite-Shot Statistics}\label{sec:circuit-bin-statistics}

\begin{figure}[t]
  \centering
  \includegraphics[width=\linewidth]{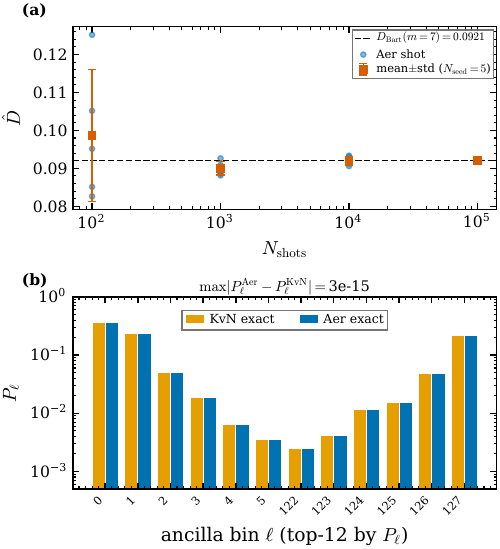}
  \caption{\textbf{Ancilla-bin distribution and finite-shot statistics from QPE circuit simulations.}
  (a) Convergence of \(\hat D\) estimated with a finite number of shots.
  The black dashed line is the state-vector value
  \(\Dbart(m_{\rm anc}=7)=0.0921\); individual points are shot-based
  simulations with independent seeds, and squares denote the mean and
  standard deviation.
  (b) Comparison of the QPE ancilla-bin probability distributions. The
  plotted bins are the 12 bins with the largest probabilities, and the
  maximum difference over all bins is
  \(\max_\ell|P^{\rm circ}_\ell-P^{\rm KvN}_\ell|\simeq3\times 10^{-15}\),
  as indicated in the figure.}
  \label{fig:circuit-qpe-agreement}
\end{figure}

Figure~\ref{fig:circuit-qpe-agreement}(a) shows the convergence of
\(\hat D\) when the same QPE circuit is executed with a finite number of
shots. Here we use \(m_{\rm anc}=7\), corresponding to \(K=2^7\). Since
\(\Delta t=0.02\),
\begin{equation}
  \tau_{\QPE}=K\Delta t=2^{m_{\rm anc}}\Delta t=2.56.
  \label{eq:tau-qpe-m7-sec6}
\end{equation}
The target value obtained from the state-vector simulation is
\(\Dbart(m_{\rm anc}=7)=0.0921\), and the mean of \(\hat D\) converges to
this value as the number of shots is increased. The standard deviation
decreases according to
\begin{equation}
  \sigma(\hat D)
  \simeq
  \frac{1}{2}\Cvv(0)\tau_{\QPE}
  \sqrt{\frac{P_0(1-P_0)}{N_{\rm shots}}}.
  \label{eq:qpe-shot-noise}
\end{equation}
This scaling follows because the bin-zero measurement is a Bernoulli
trial.

Figure~\ref{fig:circuit-qpe-agreement}(b) shows the state-vector
simulation of the QPE circuit. The ancilla-bin distribution obtained from
the discrete KvN calculation and that obtained from the Qiskit
state-vector simulation agree over all bins to within machine precision.
This confirms that the full QPE circuit, including the controlled-
\(U^{2^j}\) gates, the inverse QFT, and the bit ordering convention,
generates the same ancilla probability distribution as the discrete KvN
state-vector calculation.

Together, these two simulations show that the circuits constructed in
Sec.~\ref{sec:circuit-implementation} reproduce both the discrete KvN
propagator and the QPE bin-zero readout at the state-vector level. The
remaining finite-shot error is not a circuit-construction error, but the
statistical error in estimating \(P_0\). This statistical error is the
target of the amplitude-estimation method discussed in
Sec.~\ref{sec:amplitude-estimation}.

\bibliographystyle{apsrev4-2}
\bibliography{refs}

\end{document}